\documentclass[aps,prd,preprint,superscriptaddress,nofootinbib]{revtex4-1}
\usepackage[utf8]{inputenc}
\usepackage{amsmath}
\usepackage{amssymb}
\usepackage{graphicx}  
\usepackage{mathtools}
\usepackage{hyperref}
\usepackage[caption=false]{subfig}

\numberwithin{equation}{section}

\newcommand{\abs}[1]{\left\lvert #1 \right\rvert}
\renewcommand{\tilde}{\widetilde} 

\newcommand{\beq}{\begin{eqnarray}}
\newcommand{\eeq}{\end{eqnarray}}

\begin{document}

\title{$a$-Anomalous Interactions of the Holographic Dilaton}

\author{Csaba Cs\'aki}
\email{csaki@cornell.edu}
\affiliation{Department of Physics, LEPP, Cornell University, Ithaca, NY 14853, USA}

\author{Jay Hubisz}
\email{jhubisz@syr.edu}
\affiliation{Department of Physics, Syracuse University, Syracuse, NY 13244, USA}

\author{Ameen Ismail}
\email{ai279@cornell.edu}
\affiliation{Department of Physics, LEPP, Cornell University, Ithaca, NY 14853, USA}

\author{Gabriele Rigo}
\email{gabriele.rigo@ipht.fr}
\affiliation{Université Paris-Saclay, CNRS, CEA, Institut de Physique Théorique, 91191, Gif-sur-Yvette, France}

\author{Francesco Sgarlata}
\email{fsgarlata@cornell.edu}
\affiliation{Department of Physics, LEPP, Cornell University, Ithaca, NY 14853, USA}

\date{\today}

\begin{abstract}
    We explore higher-derivative terms in the low-energy effective action for the dilaton, the Goldstone boson of spontaneously broken scale invariance. Focusing on the simplest holographic realization of spontaneously broken scale invariance, the Randall--Sundrum (RS) scenario, we identify the nonlinear action for the RS dilaton by integrating out Kaluza--Klein graviton modes.
    The coefficient of a particular four-derivative dilaton self-interaction can be identified with the Weyl $a$-anomaly of the dual conformal field theory, which we use to verify anomaly matching arguments. We also find novel, $a$-dependent couplings of the dilaton to light matter fields.
    These anomalous interactions can have a significant effect on the collider phenomenology and the cosmology, potentially allowing us to probe the structure of the underlying conformal sector via low-energy physics.
    The dilaton effective theory also serves as an interesting scalar analog of gravity, and we study solutions to the equation of motion that parallel black holes and cosmologies.
\end{abstract}

\maketitle

\section{Introduction}\label{sec:intro}

Many extensions of the Standard Model (SM) involve a spontaneously broken conformal sector. These include holographic composite Higgs models~\cite{Agashe:2004rs} (see~\cite{Contino:2010rs,Bellazzini:2014yua,Panico:2015jxa} for reviews), dark matter models from a hidden conformal sector~\cite{Gherghetta:2010cq,vonHarling:2012sz,Hong:2019nwd,Chaffey:2021tmj}, continuum dark matter~\cite{Csaki:2021gfm,Csaki:2021xpy} and continuum naturalness~\cite{Csaki:2018kxb}, crunching solutions to the cosmological constant problem or hierarchy problem~\cite{Bloch:2019bvc,Csaki:2020zqz}, etc. One of the most important characteristics of such models is the appearance of a dilaton-like particle, the Goldstone boson associated with broken scale invariance, which would clearly signal the presence of the underlying conformal sector.\footnote{For our purposes, we will only be concerned with scale-invariant theories that are also conformally invariant. Hence, we have no need to distinguish scale and conformal symmetries. See~\cite{Dymarsky:2013pqa,Nakayama:2013is} for discussion of what conditions are necessary for scale invariance to imply conformal invariance.} Understanding the detailed properties of such a dilaton could therefore play an important role in identifying the properties of the conformal sector.

In this work, we focus on the role that anomalies play in shaping the interactions of the low-energy dilaton theory.  There is historical precedence:  in the chiral Lagrangian, decay of the neutral pion mainly proceeds through anomalous violation of the $U(1)_A$ symmetry.  Computation of the anomaly  predicted a decay width of $\Gamma(\pi_0 \rightarrow \gamma \gamma) = \frac{\alpha^2}{576 \pi} \frac{m_\pi^3}{f_\pi^2} N_C^2$.  This provided an experimental probe of an aspect of the UV theory, the number of QCD colors, from IR physics.  The anomaly also shed light on the $\eta$-$\eta'$ puzzle. We are interested in whether there are similar low-energy probes of near-conformal sectors, where interactions of the dilaton offer insight into physics of the high-energy conformal sector.

In holographic duals based on Randall--Sundrum (RS)-type~\cite{Randall:1999ee} warped extra dimensions, the dilaton is identified with the radion field~\cite{Goldberger:1999uk,Csaki:1999mp,Goldberger:1999un}. The radion is massless in the original RS model, corresponding to purely spontaneous symmetry breaking. The leading properties of the radion were discussed in~\cite{Csaki:1999mp,Goldberger:1999un,Csaki:2000zn,Csaki:2007ns}, while the holographic interpretation was given in~\cite{Rattazzi:2000hs,Arkani-Hamed:2000ijo}.
Most of the experimentally relevant couplings of the dilaton/radion to other particles follow from a spurion analysis, and lead to verifiable predictions for the various couplings~\cite{Bellazzini:2012vz,Chacko:2012sy}. These are the so-called ``low-energy theorems" for the dilaton.  However, the pure dilaton Lagrangian is less trivial: it involves a tree-level quartic, which, unlike for other Goldstone bosons, is not forbidden by the global symmetry~\cite{Fubini:1976jm,CPRtalk1,*CPRtalk2,Bellazzini:2013fga}. This would generically destabilize the dilaton potential, and hence a realistic model can only be obtained if another potential term---corresponding to some explicit breaking of the scale symmetry---balances the quartic, giving rise to a finite dilaton vev and positive mass squared (e.g.~the Goldberger--Wise mechanism~\cite{Goldberger:1999uk}). 

The aim of this paper is to start exploring the higher-derivative terms in the dilaton action, and to understand what information they contain about the underlying conformal field theory (CFT). 
We focus on the lowest-order term allowed by scale invariance 
\begin{equation}
{\mathcal L}_\text{anomaly} = 2 a (\partial \tau )^4
\end{equation}
where $\tau$ is the dimensionless dilaton. The coefficient $a$ of this operator is closely related to the famous $a$-anomaly featured in the $a$-theorem of Komargodski and Schwimmer~\cite{Komargodski:2011vj}. The $a$-anomaly has been studied extensively in unbroken CFTs; this has been especially productive in the case of supersymmetric CFTs and in holography~\cite{Henningson:1998gx,Nojiri:1999mh,Freedman:1999gp,Girardello:1999bd,Imbimbo:1999bj,Myers:2010tj}. We will argue that the $(\partial \tau)^4$ term, and thus the anomaly, can actually be calculated for any given holographic model of spontaneously broken scale invariance. Hence, measuring this coupling could provide additional important information about the structure of the underlying CFT. We concentrate on the RS scenario, the simplest holographic model of a spontaneously broken CFT. To our knowledge, the existence of the $(\partial \tau)^4$ term and other anomalous interactions in such models, although quite evident, has not been appreciated in previous studies.

The paper is organized as follows. We begin by reviewing some aspects of dilaton effective actions in \S~\ref{sec:effdilaton}, emphasizing the relationship between the dilaton and the $a$-anomaly. This sets the stage for the holographic computation of the dilaton effective action and the $a$-anomaly in the RS scenario in \S~\ref{sec:holdilaton} (we defer calculational details to Appendix~\ref{sec:RScalculations}). We find
\begin{equation}
    a_{\rm RS} = \frac{1}{8\kappa^2k^3},
\end{equation}
where $\kappa$ is related to the 5D Planck scale via $1/(2\kappa^2) = M_5^3$, and $k$ is the inverse AdS curvature. Comparing this result to the $a$-anomaly in the unbroken phase of the theory, we verify anomaly matching arguments. We discuss corrections to our result from explicit breaking effects and higher curvature terms (with detailed computations in Appendix~\ref{sec:GBgravity}), the latter encoding subleading terms in the $1/N$ expansion.
In addition to the $(\partial \tau)^4$ term, we also find $a$-dependent couplings of the dilaton to matter. In \S~\ref{sec:pheno} we comment on the resulting implications of the $a$-anomalous interactions for collider phenomenology and cosmology.
The dilaton also serves as an interesting scalar analog of gravity, which we explore in \S~\ref{sec:scalargravity}.

\section{Effective theory of the dilaton}\label{sec:effdilaton}

When spacetime symmetries are spontaneously broken, there are Goldstone bosons associated to the broken symmetry generators. The dilaton is the sole Goldstone boson of spontaneously broken conformal symmetry\footnote{There is only one Goldstone boson, as the number of Goldstones can be smaller than the number of broken generators, unlike the breaking of internal symmetries. This is due to a redundancy in parametrizing fluctuations of the order parameter with elements of the coset~\cite{Low:2001bw,Volkov:1973vd}.}.

There is a useful method to write effective actions for the dilaton $\tau$~\cite{Komargodski:2011vj}, which we review here. We define the effective metric $\tilde{g}_{\mu\nu} = e^{-2\tau} g_{\mu\nu}$, where $g_{\mu\nu}$ is the background metric. Under a Weyl transformation, $g_{\mu\nu} \rightarrow e^{2\alpha} g_{\mu\nu}$ and the dilaton shifts as $\tau \rightarrow \tau + \alpha$. Then any diffeomorphism-invariant action constructed from $\tilde{g}$ is automatically conformally invariant. Building an action in this way is equivalent to coset construction methods. The simplest action we can write using this method  is just the Einstein--Hilbert action:
\begin{equation}\label{eq:4Deinsteinhilbert}
    S = \frac{f^2}{12} \int d^4 x \sqrt{\abs{\tilde{g}}} \left( \tilde{R} + 2 \Lambda \right)
\end{equation}
where $\tilde{R}$ denotes the Ricci scalar constructed from the effective metric $\tilde{g}$, and $f$ may be associated to the scale of spontaneous symmetry breaking.
In a Minkowski background $g_{\mu\nu} = \eta_{\mu\nu}$, Eq.~\eqref{eq:4Deinsteinhilbert} is given in terms of $\tau$ by 
\begin{equation}\label{eq:einsteinhilbertexpanded}
    S = \frac{f^2}{12} \int d^4 x \left[ 6  e^{-2\tau} (\partial \tau)^2 + 2 \Lambda e^{-4 \tau} \right] .
\end{equation}
This can be put into a more familiar form by a field redefinition $\phi = f e^{-\tau}$, setting $\Lambda =\lambda/4f^2$, and then expanding $\phi$ about a symmetry-breaking vev $\phi = f - \varphi$, leading to
\begin{equation}
    S = \int d^4 x \left[ \frac{1}{2} (\partial \varphi)^2 +\frac{\lambda}{24} f^4 \left(1- \frac{\varphi}{f}\right)^4 \right] .
\end{equation}
These are just the usual dilaton kinetic and quartic terms permitted by scale invariance.

$\lambda > 0$ would drive $f$ to $0$, corresponding to $\tau \rightarrow \infty$; $\lambda < 0$ would lead to a runaway potential for $f$. These scenarios can only be stabilized by an explicit breaking of scale invariance, as for example in the Goldberger--Wise mechanism~\cite{Goldberger:1999uk}.   In this case, non-quartic potential terms can fix the breaking scale $f$.
Hence, if the conformal symmetry is truly spontaneously broken, the ``cosmological constant'' term in Eq.~(\ref{eq:4Deinsteinhilbert}) must vanish. This leads to the equation of motion $(\partial \tau)^2 = \Box \tau$ (or equivalently, $\Box \phi = 0$), up to higher-derivative terms.

Terms with more derivatives arise from introducing higher-order curvature terms to Eq.~\eqref{eq:4Deinsteinhilbert}, such as $\tilde{R}^2$. We note, however, that all the possible four-derivative terms that can be constructed in this way vanish on the two-derivative equation of motion.

Next we consider the anomaly terms.
In general, a field theory that enjoys conformal symmetry in a flat spacetime background does not preserve its conformal invariance in a curved background. At the quantum level this gives rise to trace anomalies, manifesting in a nonvanishing expectation value for the trace of the stress-energy tensor. In four dimensions, the most general form of the anomaly is~\cite{Duff:1993wm}
\begin{equation}\label{eq:generalanomaly}
    \langle T^\mu_\mu \rangle = c W_{\mu\nu\rho\sigma}^2 - a \left( R_{\mu\nu\rho\sigma}^2 - 4 R_{\mu\nu}^2 + R^2 \right) - a' \nabla^2 R .
\end{equation}
The $c$-anomaly term is proportional to the square of the Weyl tensor $W_{\mu\nu\rho\sigma}$, and the $a$-anomaly term is proportional to the Euler density. The $a'$ term is not of interest because it can be removed by adding a finite, local counterterm to the action.

The $a$-anomaly manifests as a term in the effective dilaton action that cannot be written in terms of $\tilde{g}_{\mu\nu}$, but is nevertheless conformally invariant. This takes the form~\cite{Schwimmer:2010za,Fradkin:1983tg,Riegert:1984kt}
\begin{equation}\begin{split}
    S_{a} &= {\tiny a \int d^4 x  \sqrt{\abs{g}} \left[- \tau \left( R_{\mu\nu\rho\sigma}^2 - 4 R_{\mu\nu}^2 + R^2 \right) - 4G^{\mu\nu} \partial_\mu \tau \partial_\nu \tau + 4 (\partial \tau)^2 \Box \tau - 2 (\partial \tau)^4 \right] } \\
    &\xrightarrow{\rm Minkowski} 2a \int d^4 x (\partial \tau)^4 \: (+ \text{ higher derivative terms}).
\end{split}\label{eq:Sanomaly}\end{equation}
The first line holds in an arbitrary curved spacetime, and one can check that its variation under a conformal transformation yields precisely the $a$-anomaly term in Eq.~\eqref{eq:generalanomaly}. In the second line we specify to a Minkowski background and use the equation of motion for $\tau$. The essential point is that the $a$-anomaly can be determined by computing the $(\partial \tau)^4$ term in the effective action.

\section{Effective Theory of the Holographic Dilaton}\label{sec:holdilaton}

In the 5D RS I geometry~\cite{Randall:1999ee}, the presence of an IR brane that ends the extra spatial dimension leads to the existence of a ``radion" mode corresponding to fluctuations of that brane. The AdS/CFT duality relates this radion to the dilaton of spontaneously broken conformal invariance.  The IR brane, essentially the vacuum expectation value (vev) of the radion, corresponds to a vev of operators in the dual CFT that spontaneously break the conformal symmetry~\cite{Rattazzi:2000hs,Arkani-Hamed:2000ijo}.

Motivated by the duality, in this Section we aim to compute the RS radion effective action directly from the 5D gravity theory and match it onto the dilaton effective theory outlined in \S~\ref{sec:effdilaton}. Working in a derivative expansion up to order $\partial^4$, we integrate out Kaluza--Klein (KK) graviton modes at tree-level, yielding a low-energy effective action for the radion. We can then read off the $a$-anomaly from the coefficient of the $(\partial \tau)^4$ term. The AdS/CFT correspondence converts the difficult problem of computing the $a$-anomaly in a strongly coupled CFT into a perturbative 5D gravitational calculation.

In previous work, the radion was identified by working perturbatively in the field fluctuations and reading off the spectrum. We will shortly see that the derivative expansion we employ is more closely connected to the procedure of integrating out KK gravitons. To our knowledge, our action at the two-derivative level comprises the first calculation of the RS radion kinetic term at all orders in the field fluctuations. We also remark that the four-derivative terms in the action correspond precisely to the anomalous dilaton interactions in Eq.~\eqref{eq:Sanomaly}, along with the corresponding dilaton-matter and matter-matter interactions.

\subsection{Setting up the holographic dilaton action}
Our starting point is a 5D gravity theory compactified on an interval $y \in [ y_0 ,  y_1 ]$ with negative bulk cosmological constant.  For now we consider only Einstein--Hilbert terms, and consider modifications of the gravity theory at the end of this Section.  The action is
\begin{equation}\label{eq:einsteinhilbert}
    S = -\frac{1}{2\kappa^2} \int d^5 x \sqrt{g} \left( R+ 2\Lambda \right) - \sum_{i=0,1}\frac{1}{\kappa^2} \int d^4 x \sqrt{h_i} \left(K_i+\lambda_i\right),
\end{equation}
with $\Lambda = -6 k^2$, where $k$ will be identified with the inverse AdS curvature. The induced metric on and extrinsic curvature of the $y_0$ (UV) and $y_1$ (IR) branes are denoted by $h_i$ and $K_i$ respectively, and give the Gibbons--Hawking--York contributions to the compactified gravity theory. The $\lambda_i$ are the tensions of the branes which terminate the geometry. Note that in our parametrization, the locations of the branes are fixed, and the radion lives entirely in the 5D metric.

We also include a simplified model for interactions with light matter ($m \lesssim f$), which we take to be completely localized on the IR brane.  More realistic scenarios involve bulk fields, but have couplings to the dilaton which differ only by order one factors from our simpler model.  The matter action is
\beq\label{eq:matteraction}
S_\text{matter} = \int d^4x \sqrt{h_1} \mathcal{L}(h_1, \{ \psi_\text{light} \} ).
\eeq

We can parametrize the 5D metric in a gauge where we fix $g_{\mu 5} = 0$. The metric then decomposes as a traceless 4D tensor, $h_{\mu\nu}$, and a scalar dilaton mode $A$:
\beq\label{eq:metricansatz}
ds^2 = e^{-2 A(x,y)} \left(\eta_{\mu\nu}+h_{\mu\nu}(x,y)\right) dx^\mu dx^\nu - B^2(x,y) dy^2.
\eeq
The function $B$ can be related to the other degrees of freedom through the extra-dimensional Einstein equations, while $h_{\mu\nu}$ parametrizes the tensor fluctuations. We will integrate out these modes to obtain the low-energy effective action for the holographic dilaton.

When we integrate out the tensor by solving its equation of motion, $h_{\mu\nu}$ will be a function of the only remaining light bulk field, the scalar mode $A$. Thus, $h_{\mu\nu}$ will contain at least two 4D derivatives of $A$ to have the right transformation properties under the 4D Lorentz group. 
Also, as long as there is no mixing between $A$ and $h_{\mu\nu}$ (which just amounts to solving the linearized Einstein equations), the action will be quadratic in $h_{\mu\nu}$.

From these arguments, it follows that the action at the two-derivative level is just the one with $h_{\mu\nu}$ set to zero. At the next order in integrating out $h_{\mu\nu}$, we obtain the four-derivative effective action. In the remainder of this Section we compute the action at $\mathcal{O}(\partial^2)$ and then at $\mathcal{O}(\partial^4)$.

\subsection{Power counting for the derivative expansion}

It is instructive to first get an expectation for the coefficients of the higher-derivative terms by doing a power counting in the 5D parameters.  From holography, and from the fact that the dimensionless $h_{\mu\nu}$ must scale as $h \sim \left(\frac{\partial A}{k}\right)^2$, we can expect that from expanding perturbatively in $h_{\mu\nu}$, the effective Lagrangian will ultimately be of the scale-invariant form
\beq\begin{split}
    {\mathcal L} &= e^{-4A_{\rm IR}} \Lambda \\
    &+\frac{k}{\kappa^2} \left[ \alpha_2 e^{-2A_{\rm IR}} \left( \frac{\partial A_{\rm IR}}{k} \right)^2 +\alpha_4  \left( \frac{\partial A_{\rm IR}}{k} \right)^4 + \alpha_6 e^{2 A_{\rm IR}}  \left( \frac{\partial A_{\rm IR}}{k} \right)^6 + \cdots \right],
\end{split} \eeq
 where $A_{\rm IR}$ is the metric function evaluated at $y_1$, and the $\alpha_i$ are $\mathcal{O}(1)$ parameters obtained by expanding the action and performing integrals over the extra dimension.  We have neglected contributions from the UV brane, which will add various conformal symmetry-breaking terms to the effective action.
We can put the action into canonical form with the field redefinition $\phi = \sqrt{ \frac{2\alpha_2}{\kappa^2 k}} e^{-A_{\rm IR}}$, yielding
\beq
{\mathcal L} = \tilde{\Lambda} \phi^4+ \frac{1}{2} (\partial \phi)^2 + \frac{\alpha_4}{\kappa^2 k^3} \frac{(\partial \phi)^4}{\phi^4} + \frac{2 \alpha_2 \alpha_6}{(\kappa^2 k^3)^2} \frac{(\partial \phi)^6}{\phi^8} + \cdots
\eeq
with each term in the series being enhanced by additional factors of the number of colors $N$ of the dual CFT.  We assume the matching relation $N^2 = 16\pi^2 / (2 \kappa^2 k^3)$~\cite{Agashe:2004rs,Contino:2010rs,vonHarling:2017yew}, derived from the exact AdS/CFT duality involving Type IIB string theory on AdS$_5 \times S^5$, holds.
Notably, the coefficient of the $a$-anomaly term is expected to be of order $\frac{N^2}{8\pi^2}$.

Finally, we write the action in perhaps a more familar form for phenomenology by expanding $\phi$ about a vev $f$, $\phi = f (1 - \varphi/f)$, and with the identification of the KK scale $M_\text{KK} = k e^{-\langle A_{\rm IR}\rangle} = \sqrt{\frac{\kappa^2 k^3}{2\alpha_2}} f$:
\beq\begin{split}
    {\mathcal L} &= \tilde{\Lambda} f^4 (1- \varphi/f)^4 \\
    &+ \frac{1}{2} (\partial \varphi)^2 + \frac{\alpha_4}{4\alpha_2^2} \frac{\kappa^2 k^3}{M_{\rm KK}^4} \frac{(\partial \varphi)^4}{(1-\varphi/f)^4} + \frac{\alpha_6}{8 \alpha_2^3} \frac{(\kappa^2 k^3)^2}{M_{\rm KK}^8}\frac{(\partial \varphi)^6}{\left( 1- \varphi/f\right)^8} + \cdots
\end{split}\eeq
Note that if the mass scale of the extra-dimensional resonances is held fixed while varying the $N$ of the dual CFT, the tower of operators is suppressed by factors of $1/N^2$.  This is due to the usual hierarchy between the dilaton decay constant and the scale of composite resonances, which grows linearly with $N$.

Here we have neglected light matter fields, with mass $\lesssim f$, that source the KK gravitons.  Including them ensures the higher-derivative terms will be accompanied by insertions of the stress-energy tensor of those fields, since via its equation of motion, we expect $h_{\mu\nu} \ni \{\frac{1}{k^2} (\partial A_{\rm IR})^2,\frac{\kappa^2}{k} T_{\mu\nu} \}$.

\subsection{$\mathcal{O}(\partial^2)$: the kinetic term}

With $h_{\mu\nu}\rightarrow 0$, the $y$-$y$ Einstein equation relates the $B$ and $A$ functions, such that at order $\partial^2$ we have 
\begin{equation}\label{eq:metricansatzsimplified}
    ds^2 = e^{-2 A} \eta_{\mu\nu} dx^\mu dx^\nu - \frac{A'^2}{k^2 - \frac{1}{2} e^{2 A} \left[\Box A - (\partial A)^2 \right]}  dy^2 .
\end{equation}
The action Eq.~\eqref{eq:einsteinhilbert} is seen to be a total $y$-derivative after expanding with this metric ansatz, allowing us to trivially perform the integral over $y$. This leads to the effective RS radion action
\begin{equation}\label{eq:effaction}
    S_\text{radion} = \frac{3}{\kappa^2 k} \int d^4x\left. e^{-2A} (\partial A)^2  \right|^{y_1}_{y_0} -\int d^4x \left[ \lambda_1 + \frac{6 k}{\kappa^2} \right]  e^{-4A_\text{IR}} - \left[ \lambda_0 -  \frac{6 k}{\kappa^2} \right] e^{-4A_\text{UV}} ,
\end{equation}
where $A_{\rm UV/IR} = A(y_{0,1})$.
We present the full details of this calculation in Appendix~\ref{sec:RScalculations}.

Two tunings, $\lambda_0 = 6k/\kappa^2$ and $\lambda_1 = -6k/\kappa^2$, set respectively the cosmological constant and the scale-invariant quartic to zero, and admit static solutions to the equations of motion.

The radion action above is \textit{not} in a conformally invariant form because the UV brane explicitly breaks scale invariance. Suppose one tunes the UV brane tension to remove the bare cosmological constant, setting $\lambda_0 = 6k/\kappa^2$. Then the potential is of the desired form $e^{-4 \tau}$,  with $\tau \equiv A_\text{IR}-A_1$ (we define $A_{0,1} = \langle A(y_{0,1}) \rangle$, the background values of $A$ on the branes), and arises completely from the IR brane term. 
However, the kinetic term is not in a manifestly scale-invariant form due to the contribution from the UV brane.  In terms of $\tau$ and $A_{0,1}$,
\beq
{\mathcal L}_\text{kin} \approx \frac{3}{\kappa^2 k} e^{-2 A_1} e^{-2\tau} (\partial \tau)^2 \left[ 1-e^{2(A_0-A_1)} e^{-2 \tau} \right],
\eeq
where we have neglected terms with additional factors of $e^{-2 (A_1 - A_0)}$. One could perform a field redefinition to put the kinetic term into the desired form, but then the other terms in the dilaton action would not be manifestly scale-invariant. This remains the case even if we were to further tune $\lambda_1$ to remove the IR quartic, since the dilaton action also includes higher-derivative terms.

We conclude that the radion action is only scale-invariant in the limit that $A_1 -A_0 \rightarrow \infty$, which corresponds to sending the UV brane to the AdS boundary.  In this case, one is left with the simpler action
\beq
\label{eq:rsradion}
S_{\rm radion} = \int d^4x \frac{1}{2} f^2 e^{-2\tau} (\partial \tau)^2 - \lambda f^4 e^{-4\tau}
\eeq
which takes the desired form with $\tau$ being the dilaton.
The dilaton decay constant $f$ is given by $f^2 = \frac{6}{\kappa^2 k} e^{-2 A_1}$.

Lastly, we identify the matter couplings by expanding the matter action Eq.~\eqref{eq:matteraction} with the metric ansatz, leading to
\beq
    S_{\rm radion} = \int d^4x \frac{1}{2} f^2 e^{-2\tau} (\partial \tau)^2 - \lambda f^4 e^{-4\tau} + \tau T^\mu_\mu
\eeq
where $T_{\mu\nu}$ is the matter stress-energy tensor. 
This is our final result for the radion action at order $\partial^2$.

Take note that the Einstein equations are not yet fully solved if the limit $h_{\mu\nu}=0$ is taken.  In particular, the $\mu$-$\nu$ equations have terms that are quadratic in derivatives, and which can only be balanced by contributions from the tensor, $h_{\mu\nu}$.  If we continue to work in the limit $A_1-A_0 \rightarrow \infty$, solving for the tensor amounts to integrating out the massive graviton KK modes.  There is no graviton zero mode since, in this conformal limit, the massless 4D graviton decouples (or equivalently becomes non-normalizable).  Including interactions of the massless graviton would generate additional violations of conformal invariance, due to introduction of the effective 4D Planck scale.

\subsection{$\mathcal{O}(\partial^4)$: the $a$-anomaly} 

To obtain the holographic action at four derivatives, we integrate out the KK gravitons at tree-level. As explained above, this amounts to solving the classical equation of motion for the 4D tensor $h_{\mu\nu}$ at second order in derivatives. We then substitute this solution back into the action, expanding to order $h_{\mu\nu}^2$ (corresponding to order $\partial^4$), and integrate out the extra dimension. Again, we present the full details of this calculation in Appendix~\ref{sec:RScalculations}.

Expanding the 5D action (Eqs.~\eqref{eq:einsteinhilbert} and \eqref{eq:matteraction}) using the metric ansatz Eq.~\eqref{eq:metricansatz} yields the following equation of motion for $h_{\mu\nu}$ at the level of two derivatives:
\begin{equation}
    \partial_y \left( \frac{e^{-4A}}{A'} h'_{\mu\nu} \right) =\frac{2}{k^2} \partial_y \left[ e^{-2A} \left( \partial_\mu \partial_\nu A + \partial_\mu A \partial_\nu A + \frac{1}{2} \eta_{\mu\nu} \left((\partial A)^2 - 2 \Box A \right) \right) \right] \equiv 2 \partial_y J_{\mu\nu} ,
\end{equation}
subject to the boundary condition
\begin{equation}
    \frac{e^{-4A}}{A'} h_{\mu\nu}' \Big |_{y_1} = \frac{3}{2k^2} e^{-2A_{\rm IR}} \left( \Box A_{\rm IR} - (\partial A_{\rm IR})^2 \right) \eta_{\mu\nu} - \frac{\kappa^2}{k} T_{\mu\nu} ,
\end{equation}
where $T_{\mu\nu}$ is the stress-energy tensor associated to the matter action.
The solution to this is
\begin{equation}
    h'_{\mu\nu} =  A' e^{4 A} \left[ 2 J_{\mu\nu}(y_1) - 2 J_{\mu\nu}(y) - \frac{\kappa^2}{k} \left( T_{\mu\nu} - \frac{1}{4} \eta_{\mu\nu} T \right) \right] .
\end{equation}

The quadratic action for $h_{\mu\nu}$ (including Einstein--Hilbert, Gibbons--Hawking--York, and brane tension terms) comes out to be
\begin{equation}
    S_{\rm tensor} = -\frac{k}{4\kappa^2} \int d^5 x \left( \frac{e^{-4A}}{A'} (h_{\mu\nu}')^2 + 4 h^{\mu\nu} \partial_y J_{\mu\nu} \right) + \frac{1}{2} \int d^4 x h_{\mu\nu} T^{\mu\nu} \Big |_{y=y_1} .
\end{equation}
Using the above solution for $h_{\mu\nu}'$ and the radion equation of motion, and integrating by parts, one can rewrite this as a total $y$-derivative. Performing the trivial integral over $y$ then leads to
\begin{equation}\label{eq:RStensoraction}\begin{split}
    S_{\rm tensor} = \frac{1}{4\kappa^2 k^3} &\int d^4 x \left[ (\partial A_\text{IR})^4 + \kappa^2 k e^{2A_{\rm IR}} \partial_\mu A_{\rm IR} \partial_\nu A_{\rm IR} \left(T^{\mu\nu} - \frac{1}{6} \eta^{\mu\nu} T \right) \vphantom{\left( T_{\mu\nu} - \frac{1}{4} \eta_{\mu\nu} T \right)^2} \right. \\
    &\left. + \frac{1}{4} \kappa^4 k^2 e^{4A_{\rm IR}} \left( T_{\mu\nu} - \frac{1}{4} \eta_{\mu\nu} T \right)^2  \right] .
\end{split}\end{equation}

With the identification $A_\text{IR} = A_1 + \tau$, and the dilaton decay constant $f^2 = \frac{6}{\kappa^2 k} e^{-2A_1}$, we can now summarize our low-energy effective action for the RS radion coupled to IR-localized matter, complete to fourth order in a derivative expansion:
\begin{equation}
\begin{aligned}
    S_\text{radion} &=  \int d^4 x  \frac{f^2}{2} e^{-2 \tau} (\partial \tau)^2  - \lambda f^4 e^{-4 \tau} + \tau T^\mu_\mu \\
    &\phantom{{}={}}+
    \frac{1}{4\kappa^2k^3} \left\{ \left[ \partial_\mu \tau  \partial_\nu \tau + \frac{3e^{2\tau}}{f^2} \left( T_{\mu\nu} - \frac{1}{4} \eta_{\mu\nu} T \right) \right]^2 + \frac{e^{2\tau}}{2f^2} (\partial \tau)^2 T \right\} .
\end{aligned}
\end{equation}
We now read off the dimensionless $a$-anomaly from the coefficient of the $(\partial \tau)^4$ term, and relate it to parameters associated with the dual CFT:
\beq
a_\text{RS} = \frac{1}{8 \kappa^2k^3} = \frac{N^2}{4 (16 \pi^2)} .
\label{eq:RSanomaly}
\eeq

We can connect this result to previous holographic calculations of the $a$-anomaly in the unbroken phase,  where the theory is dual to the full AdS spacetime without the IR brane. The $a$-anomaly has been long known in this scenario~\cite{Henningson:1998gx,Freedman:1999gp} and, as expected from anomaly matching arguments~\cite{Schwimmer:2010za}, equals our result~\eqref{eq:RSanomaly}.

\subsection{Explicit breaking and $1/N$ corrections}

We emphasize again that there are subleading, conformal symmetry-breaking corrections to our result. These corrections could arise from explicit symmetry breaking by a UV brane, or by bulk effects that deform the AdS geometry, like a kink over which the bulk curvature evolves from $k' \rightarrow k$.  These effects are suppressed by powers of $e^{A_0-A_1}$, where $A_0$ is taken to correspond to the warp factor at the UV brane (or wherever the explicit breaking occurs).  This is as expected---symmetry-breaking effects in the UV from heavy modes at the scale $M$ manifest as higher-dimensional operators suppressed by powers of $M$.  Those terms in the dilaton effective action are suppressed relative to the conformally invariant ones by powers of $f/M \approx e^{A_0-A_1}$. An important conclusion here is that the anomaly term in the radion effective theory is a function of the AdS curvature \emph{in the IR region of the space-time}.\footnote{If explicit conformal symmetry breaking occurs instead over a range of energy scales due to slow running of near-marginal operators, the story may be quite different. Symmetry-breaking operators will show up in the effective theory with factors of $\log A_1 / A_0$. An example of this occurs in the Goldberger--Wise stabilization mechanism.}

The anomaly coefficient in Eq.~\eqref{eq:RSanomaly} is the difference between the $a$-anomaly in the conformal unbroken phase of the theory and in the EFT where conformal invariance is realized nonlinearly by the dilaton. The latter receives contributions from all the particles in the effective theory (namely, the light IR-localized matter and the dilaton), which are $\mathcal{O}(1)$ in the $1/N$ expansion.

The dilaton effective action is also affected by 5D gravity loop effects and the contributions of higher-dimensional operators that serve as the counter-terms in their renormalization. So long as $M_5/k$ (and thus the $N$ of the CFT) is large, these effects will be subleading in power counting. To see this in more detail, consider adding a four-derivative curvature term to the bulk action. In Appendix~\ref{sec:GBgravity}, we argue that a generic four-derivative term can be rewritten as a Gauss--Bonnet term using field redefinitions. Hence we add to the bulk action
\begin{equation}
    S_{\rm GB,bulk} = \frac{\lambda_{\rm GB}}{2\kappa^2 k^2} \left[ \frac{2\kappa^2 k^3}{24\pi^3} \right]^{2/3} \int d^5 x \sqrt{g} \left( R^2 - 4 R_{ab}^2 + R_{abcd}^2 \right)
\end{equation}
where $\lambda_{\rm GB}$ is a dimensionless constant. The suppression factor of
\begin{equation}
    \left( \frac{2\kappa^2 k^3}{24\pi^3} \right)^{2/3} = \left( \frac{2}{3\pi N^2} \right)^{2/3}
\end{equation}
arises from na\"ive dimensional analysis with 5D cutoff scale $\Lambda_5^3 = 24\pi^3/(2\kappa^2)$. In Appendix~\ref{sec:GBgravity} we compute the resulting modification to the $a$-anomaly, finding
\begin{equation}
    a_{\rm GB} = a_{\rm RS} \left( 1 - 12 \lambda_{\rm GB}\left[ \frac{2\kappa^2 k^3}{24\pi^3} \right]^{2/3} \right)
\end{equation}
This correction from the Gauss--Bonnet term is manifestly subleading in the $1/N$ expansion. It is also in agreement with holographic $a$-anomaly calculations in the unbroken phase, consistent with anomaly matching~\cite{Myers:2010tj,Myers:2010jv}.

\section{Phenomenology}\label{sec:pheno}

We now comment on the potential for the $a$-anomaly term to contribute nontrivially to both collider physics and cosmology.
For these purposes, it is helpful to see the radion action in a form more convenient for phenomenology, where we have performed a field redefinition $\phi = f e^{-\tau}$, and expressed $f$ in terms of the KK mass scale, $f^2 = \frac{6}{\kappa^2 k^3} M^2_\text{KK}$:
\begin{equation}\label{eq:actionpheno}\begin{split}
    S_\text{radion} =  \int d^4 x  &\frac{1}{2} (\partial \phi)^2  - \lambda \phi^4 - \log \left( \frac{\phi}{f} \right) T^\mu_\mu  \\
    &+ \frac{\kappa^2 k^3}{144 M_{\rm KK}^4} \left( \frac{f}{\phi} \right)^4 \left\{ \left[ \partial_\mu \phi  \partial_\nu \phi + 3 \left( T_{\mu\nu} - \frac{1}{4} \eta_{\mu\nu} T \right) \right]^2 + \frac{1}{2} (\partial \phi)^2 T \right\} .
\end{split}\end{equation}
If we were to expand around a radion vev, $\phi = f - \varphi$, the $a$-anomaly term becomes a series of dimension-8 and higher operators in the 4D effective theory (valid to scale $\sim 4\pi f$).  There are also new interactions with light composite matter, suppressed by the dilaton decay constant, along with new self-interactions of the matter fields. 

\subsection{Collider probes of the $a$-anomaly}

We introduced this work with a description of how the anomalous decay of the neutral pion gave information on the number of colors in QCD.  We now have a similar answer to what information about the UV theory we may glean from measuring the anomalous four-radion interaction. Measuring $a$ gives access to the number of colors $N$ in the CFT; importantly, if $a$ is large, it signifies that the CFT has a weakly-coupled 5D gravitational description.

The four-dilaton interaction would be difficult to measure, but thankfully the $a$-anomaly also manifests in couplings of the dilaton to light matter fields. Expanding Eq.~\eqref{eq:actionpheno} around a dilaton vev $\phi = f - \varphi$, we can read off the leading (dimension-8) dilaton-matter couplings
\begin{equation}\label{eq:mattercouplings}
    \frac{\kappa^2 k^3}{24 M_{\rm KK}^4} \partial^\mu \varphi \partial^\nu \varphi \left( T_{\mu\nu} - \frac{1}{6} \eta_{\mu\nu} T \right) = \frac{\pi^2}{3 N^2 M_{\rm KK}^4} \partial^\mu \varphi \partial^\nu \varphi \left( T_{\mu\nu} - \frac{1}{6} \eta_{\mu\nu} T \right) .
\end{equation}
In principle, one can probe the value of $N$ by measuring the value of this coupling from, for example, radion production cross-sections at colliders.

We remark that the dilaton already couples to the trace of the stress-energy tensor via the dimension-five operator $\varphi T$. Hence, the $(\partial \varphi)^2 T$ coupling occurring in Eq.~\eqref{eq:mattercouplings} is not especially novel. However, the $\partial_\mu \varphi \partial_\nu \varphi T^{\mu\nu}$ coupling is very interesting. This term induces an $a$-dependent contact interaction between the radion and matter fields, including classically scale-invariant fields such as the gluon and photon. In contrast, the dimension-five coupling $\varphi T$ only generates couplings to classically scale-invariant operators at loop level via the trace anomaly, since their $T^\mu_\mu$ vanishes at tree level. This contact interaction could be important for, say, radion production via gluon fusion at hadron colliders. We leave a detailed study of such effects for future work.
    
\subsection{Cosmology of a rolling dilaton}

Next we consider the cosmology of a Universe dominated by the dilaton. While this is not a fully realistic model, it is ideal for showing the important effects that the $a$-term can have.  Here we will be agnostic about the UV completion, and we do not assume we are working in the context of the holographic dilaton, but rather with the dilaton of a generic spontaneously broken CFT.  Particularly, we are interested in models where the anomaly term dominates over some range of field configurations, and its coefficient is enhanced relative to other higher-dimensional operators in the dilaton EFT.
  
\begin{figure}
    \subfloat{%
        \includegraphics[width=0.49\textwidth]{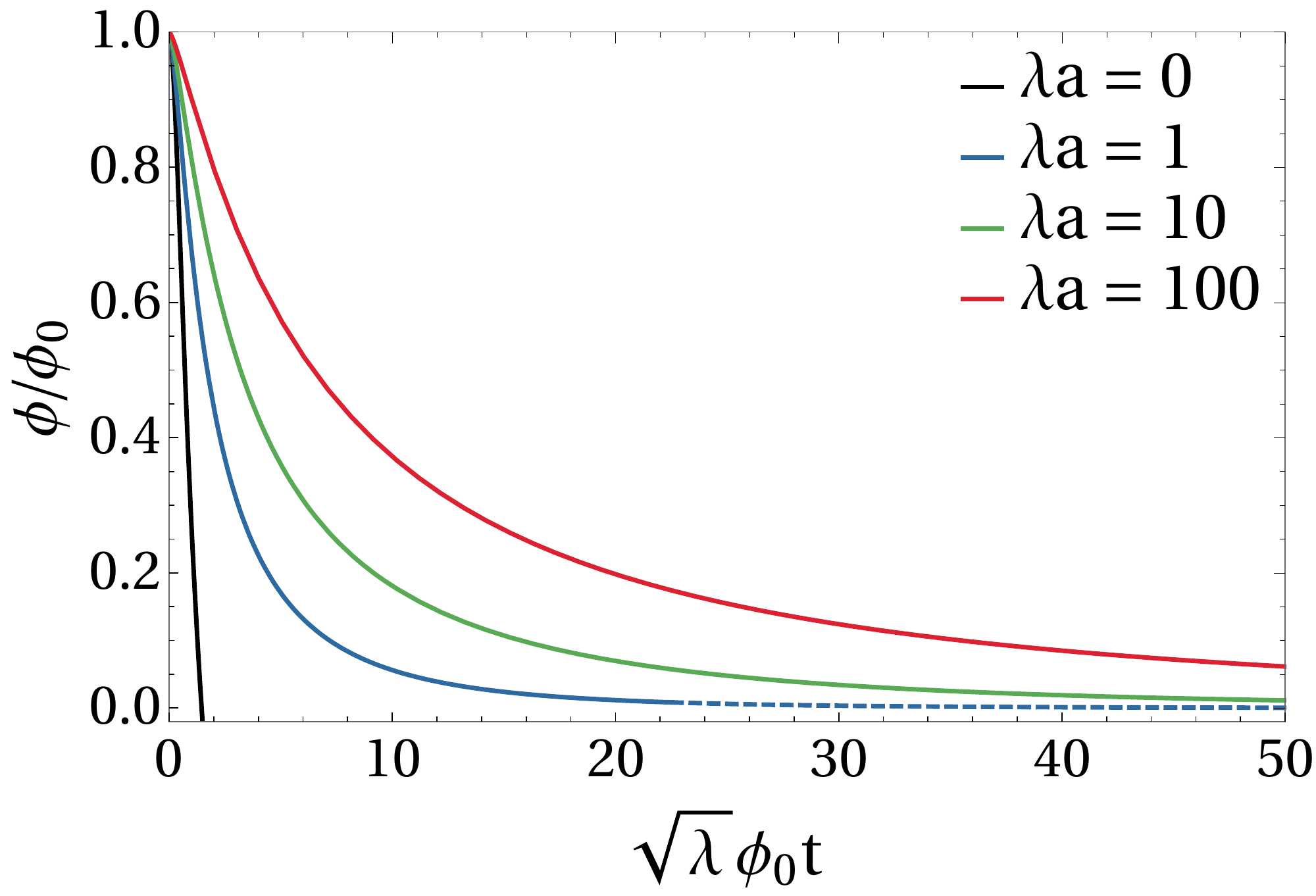}%
    }\hfill
    \subfloat{%
        \includegraphics[width=0.49\textwidth]{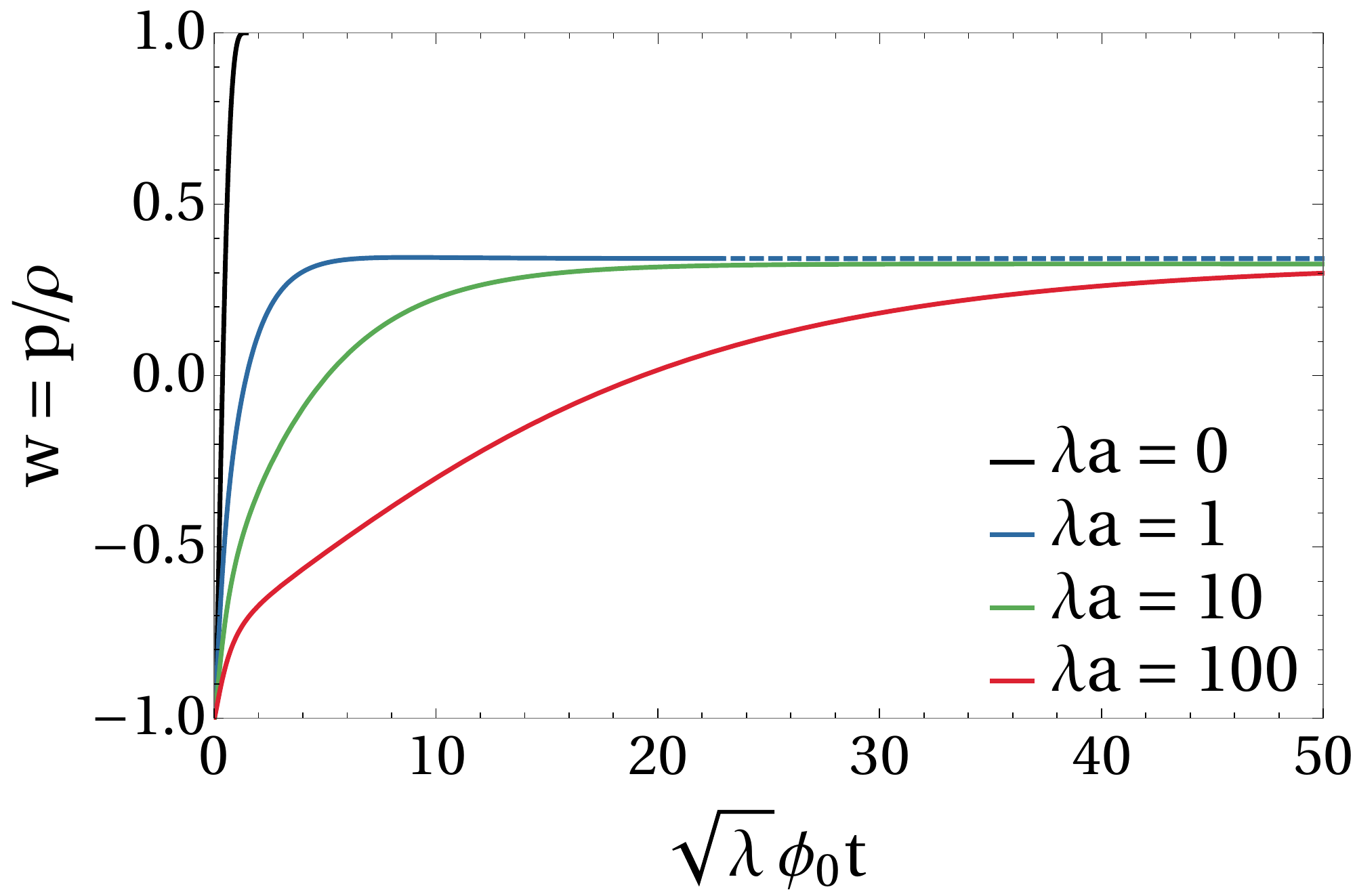}%
    }
    \caption{Time evolution (left) and equation of state parameter (right) of a spatially homogeneous dilaton field $\phi(t)$ with a positive quartic term, subject to the initial conditions $\phi_0 \equiv \phi(0) = M_{\rm Pl}$ and $\dot{\phi}(0) = 0$. When $a = 0$ (black curve), the dilaton field reaches the origin in finite time. With any nonzero anomaly term (blue, green, and red curves), $\phi$ never reaches $0$, and the equation of state approaches $w = 1/3$. The dashing in the blue curve indicates where the effect of higher-derivative terms would be important in the na\"ive dimensional analysis limit, assuming $\lambda=1$.}
    \label{fig:anomalydrag}
\end{figure}

Assuming a flat FRW metric, we show numerical solutions to the dilaton equation of motion in Fig.~\ref{fig:anomalydrag} when the dilaton quartic is taken to be positive. A Hubble friction term is generated by the coupling to gravity. The $a$-anomaly term, $2a (\partial \phi)^4 / \phi^4$, acts like a field-dependent viscosity term, generating an ``anomaly drag'' that becomes important at small values of $\phi$.

In the absence of the anomaly term, the 4D Einstein equations and dilaton equation of motion lead to
\begin{equation}
\label{eq:dilatonFRW}
\ddot{\phi} + \frac{3\dot{\phi}}{\sqrt{6}M_\text{Pl}}\sqrt{\dot{\phi}^2+2\lambda \phi^4}+4\lambda \phi^3=0\,.
\end{equation}
The dilaton rolls from its initial condition $\phi(0)=\phi_0$, $\dot{\phi}(0)=0$ down to the origin $\phi = 0$ over a time of order $t_*/\sqrt{\lambda}$, where $t_*$ depends on $\phi_0$ and $t_*\rightarrow \infty$ as $\phi_0\rightarrow 0,\infty$ (the parametric dependence on $\phi_0$ will be unimportant for the following discussion).
At $t\sim t_*/\sqrt{\lambda}$ unbroken conformal symmetry is restored, and in principle we must specify details of the full CFT (or its 5D dual) to take the story further; $\phi \rightarrow 0$ is a singularity of the EFT.

To see this in  more detail, note that near $\phi = 0$, where the field slowly changes, we can approximate $\dot{\phi}(t) \sim \frac{1}{ t}$ which, together with Eq.~\eqref{eq:dilatonFRW}, gives $\phi(t) \sim -\sqrt{2/3}M_\text{Pl}\log(t \sqrt{\lambda}/t_*)$. It is clear that derivatives do not vanish at finite times, while the field does vanish logarithmically. This means that higher-derivative operators (possibly including those that explicitly break scale invariance) might be more important than the terms retained in Eq.~\eqref{eq:actionpheno}, which signals a breakdown of the EFT.

Adding a nonzero value of $a$ dramatically changes this behavior, as shown in Fig.~\ref{fig:anomalydrag}. It prevents $\phi$ from reaching the origin in finite time, thereby smoothing out the singularity and giving the cosmology a soft exit.

Note that from na\"ive dimensional analysis with the 5D cutoff scale $\Lambda_5^3 = 24\pi^3/(2\kappa^2)$, we expect the next higher-derivative term in the action to be at least of the order
\begin{equation}
    \frac{1}{\pi^4} \left( \frac{2a^2}{3} \right)^{1/3} \frac{(\partial \phi)^6}{\phi^8} .
\end{equation}
Our calculation can no longer be trusted once this is comparable to the $a$-anomaly term. We have indicated this by dashed lines in Fig.~\ref{fig:anomalydrag}, assuming a quartic $\lambda = 1$.

Fig.~\ref{fig:anomalydrag} also shows the evolution of the dilaton equation of state parameter $w = p/\rho$, given by
\begin{equation}
    w = \frac{\dot{\phi}^2/2 - \lambda \phi^4 + 2a \dot{\phi}^4/\phi^4}{\dot{\phi}^2/2 + \lambda \phi^4 + 6a \dot{\phi}^4/\phi^4}
\end{equation}
where dots denote time derivatives. At early times, the equation of state is inflationary, with $w \approx -1$.  In the presence of a nonzero $a$-term, the dilaton evolves towards $w = 1/3$, corresponding to pure radiation, in contrast to the $w \rightarrow 1$ behavior when $a = 0$. Eventually, we expect that higher-derivative terms will decrease $w$ from $1/3$ to $0$.

We remark that if the dilaton is reponsible for inflation, the $a$-anomaly naturally provides a graceful exit in the form of a smooth transition from $w = -1$ at early times to $w = 1/3$ at late times. However, the number of inflationary e-folds for the parameters chosen in Fig.~\ref{fig:anomalydrag} range from $\sim 1$ for the $\lambda a = 1$ curve to $\sim 10$ for the $\lambda a = 100$ curve. Super-Planckian field values of order $\phi_0 \approx 10 M_P$ are needed to generate the required amount of e-folds, which is typical of quartic inflation.  We cannot expect the effective theory of spontaneously broken scale invariance to be valid above an explicit symmetry-breaking scale (in the 5D picture this would correspond to the IR brane having crossed past the UV brane).

We have emphasized above why this is a toy example: the cosmological history does not correspond to our own, and we have ignored the effects of higher-derivative and explicit symmetry-breaking terms. Nevertheless, our discussion here makes it clear that the $a$-anomaly and other higher-derivative terms may lead to a qualitative change in the time evolution of the dilaton field, with potentially interesting ramifications for cosmology.

In our example we have focused on a situation where the dilaton dominates the energy density of the universe. Another interesting direction, which is beyond the scope of this paper, is to explore the role of the anomaly term in a universe where some other field drives the cosmological expansion. This would amount to studying the dilaton equation of motion in an FLRW metric, carefully including the dilaton-matter couplings associated with the $a$-anomaly term.

\section{The $a$-anomaly and dilaton ``gravity"}\label{sec:scalargravity}

As a scalar field which couples to the stress-energy tensor, the dilaton provides an analog to gravity with solutions that parallel those in the Einstein theory. This analogy was first explored in detail in~\cite{Sundrum:2003yt}, which contains relevant background on the topic. The analogy is a good one since the dilaton EFT in 4D flat space is derived by compensating the Minkowski metric with the dilaton: $g_{\mu\nu} = (\phi/f)^2 \eta_{\mu\nu}$.  Conformally flat solutions to 4D gravity have analogous solutions in theories with spontaneously broken conformal invariance; for example there are dilaton cosmologies and dilaton ``black holes''.  This model of scalar ``gravity'' can be thought of as a simpler laboratory in which to study puzzles associated with its Einstein counterpart.

Nonrenormalizable terms in the dilaton effective action reflect properties of the underlying UV completion, hinting at how such puzzles may be resolved in dilaton gravity. For this reason, in this Section we consider the role of the $a$-anomaly term in dilatonic black holes and cosmologies. We continue to remain agnostic about the structure of higher-derivative terms.

Ignoring matter couplings now, we can express the dilaton Lagrangian as
\begin{equation}
    {\mathcal L}_\text{dilaton}= \frac{1}{2} (\partial \phi)^2 + 2a\frac{(\partial \phi)^4}{\phi^4}- \lambda \phi^4.
\end{equation}
Evolution of the dilaton in time or space can drive the dilaton towards the EFT singularity at $\phi = 0$, where higher-derivative terms like the $a$-anomaly interaction become important.

\subsection{Dilatonic ``black holes''}

Spherically symmetric solutions to the dilaton equation of motion have a structure that mimics aspects of black holes in Einstein gravity.  In the absence of the $a$-anomaly term (and with $\lambda = 0$) the dilaton profile obeys the Laplace equation for a free scalar, and a point mass creates a dilaton field profile given by 
\beq
\phi = f -\frac{\alpha}{r}.
\eeq
$f$ is the background vev of the dilaton far from the point source, and $\alpha$ is proportional to the mass of the source.
There is an EFT singularity at $r_h = \alpha/f$, where $\phi = 0$.  This is somewhat analogous to a black hole with Schwarzschild radius $r_h$. In fact, holographically this solution would correspond to a 5D black hole with some extra-dimensional profile localized on the IR brane. The full 4D CFT (or a 5D holographic dual) is presumably needed to describe the interior structure beyond the horizon.

Braneworld black holes have received a large amount of attention in the literature, and there are no known spherically symmetric, static solutions to the 5D Einstein equations in the model with only an IR brane~\cite{Chamblin:1999by,Emparan:1999wa,Emparan:1999fd,Giddings:2000ay,Emparan:2002px,Meade:2007sz}. It is worth revisiting this in the context of the dilaton effective theory with the higher-dimensional operators taken into account.

Inclusion of the $a$-anomaly interaction modifies the equation of motion, and spherically symmetric solutions obey (where primes denote derivatives with respect to $r$)
\beq
\phi'' \left(1-24 a \frac{\phi'^2}{\phi^4} \right) +\frac{2}{r} \phi' \left( 1- 8 a \frac{\phi'^2}{\phi^4}\right) + 24 a \frac{\phi'^4}{\phi^5}-4 \lambda \phi^3 = 0.
\eeq
When $\phi$ becomes small, the equation hits singularities \emph{in the approach} to the would-be horizon due to the minus signs associated with the anomaly terms in parentheses.  This happens precisely when other higher-derivative terms will also be playing a significant role, so the analysis here is not conclusive---the complete tower of operators arising from integrating out the UV CFT perhaps resolves this issue.  However, it is curious to note that the $a$-anomaly term can generate a kinetic instability in a spatially varying dilaton background in this same region of field space.  Expanding $\phi$ as $\Phi(r) + \varphi(r,t)$, the $a$-term contributes to the Lagrangian the term 
\beq
2 a \frac{(\partial \phi)^4}{\phi^4} \ni - 4a \dot{\varphi}^2 \frac{\Phi'^2}{\Phi^4}
\eeq
which can dominate over the usual kinetic term near the dilaton horizon.

Both of these issues occur when the anomaly term begins to dominate over the kinetic term.  Feeding in the $a=0$ solution to the $a$-anomaly term, we find that this occurs at a radius $r_* \sim r_h + \left( a \frac{r_h^2}{f^2} \right)^{1/4}$, signifying a breakdown of the EFT in the approach to the horizon.

So either the full CFT dynamics is needed to resolve the issue, or the kinetic instability persists, with the $a$-term giving the first hint that there are no time-independent ``black hole'' solutions to the dilaton equation of motion.  In the context of black hole studies of the 5D dual picture, there is indication that the latter explanation may be the correct one~\cite{Emparan:1999wa}.

\subsection{A dilatonic ``universe"}

We now study a scalar analog of cosmology where we associate $\phi$ with the cosmological scale factor in conformal coordinates. In this context, the dilaton quartic is akin to a cosmological constant. Without the $a$-anomaly the ``geometry'', obtained by solving the scalar equation of motion, has dS$_4$ (for $\lambda < 0$) or  AdS$_4$ (for $\lambda > 0$) solutions.

We first note that inclusion of the $a$-anomaly term still admits dS$_4$/AdS$_4$ solutions (as guaranteed by scale invariance), although with a different, $a$-dependent curvature scale:
\beq
\phi = 
\left\{ \begin{array}{ll} 
\sqrt{ \frac{ 1 + \sqrt{1+ 96 a (-\lambda)}}{4 (-\lambda)}}\frac{1}{t} & \lambda < 0 \\
\sqrt{ \frac{ 1 + \sqrt{1+ 96 a \lambda}}{4 \lambda}}\frac{1}{z} & \lambda > 0
\end{array} \right. ,
\eeq
where $t$ and $z$ are 4D time and one of the spatial coordinates, respectively.
We note that for a large $a$-anomaly, corresponding to large $N$, the curvature scales differently with the quartic, going like $(|\lambda |/a)^{1/4}$ rather than like $\sqrt{|\lambda |}$.  An enhanced $a$-anomaly appears to suppress the geometric effects of the cosmological constant term.

\begin{figure}[t!]
    \subfloat{%
        \includegraphics[width=0.49\textwidth]{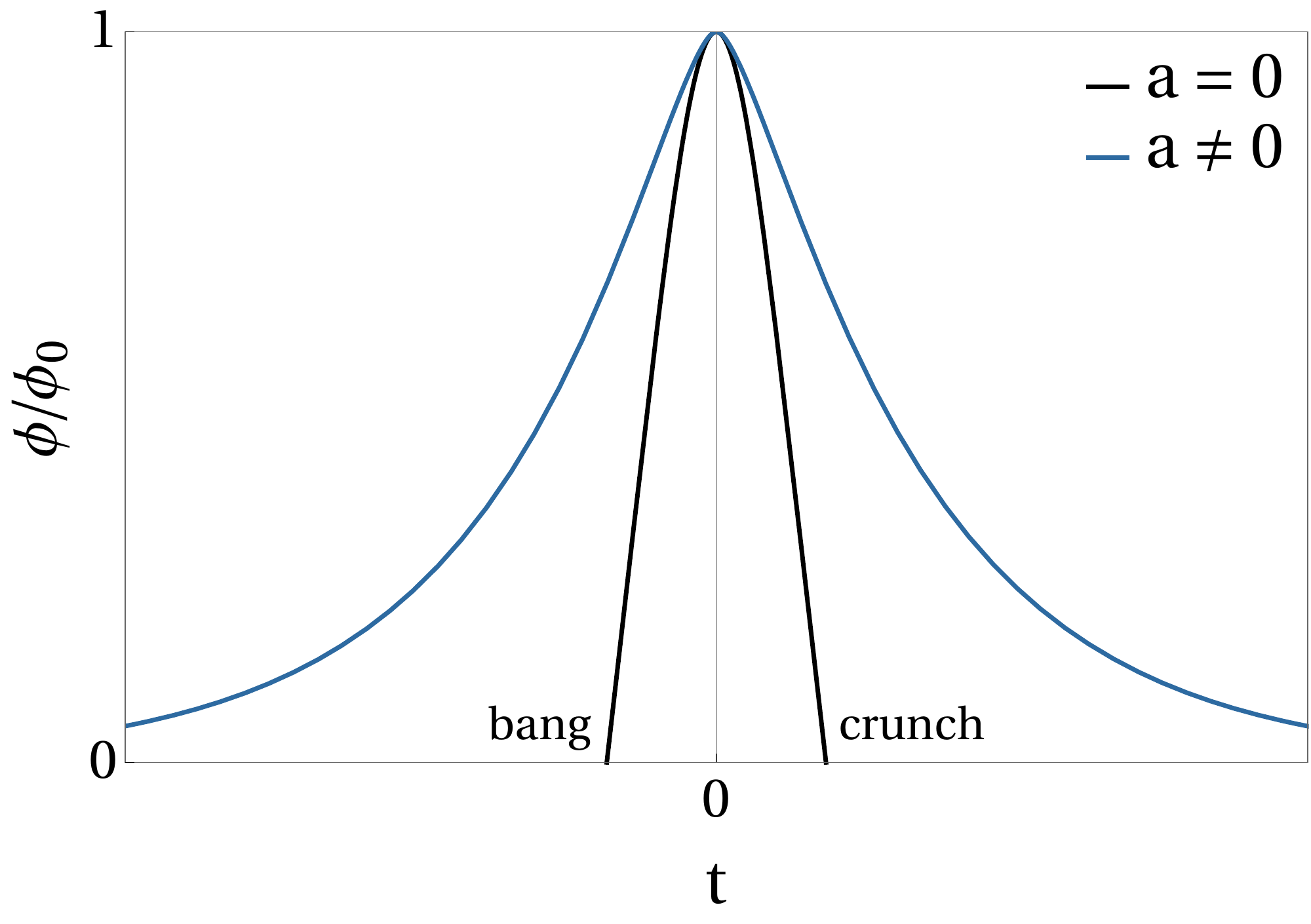}%
    }\hfill
        \caption{A cartoon of a dilaton cosmology with positive quartic. The $a$-anomaly interaction with positive coefficient renders the solution sensible in the context of the EFT.}
    \label{fig:scalarcosmo}
\end{figure}

In the absence of the $a$-term, the dilaton equation of motion admits oscillatory solutions for $\lambda > 0$. But since the EFT breaks down at $\phi = 0$, the solution is a half-cycle of this oscillation. There is a ``big bang'' following by a ``big crunch'', with the UV completion describing the dynamics before and after these singularities. We show a cartoon of this in Figure~\ref{fig:scalarcosmo}, with the black curve characterizing solutions with $a=0$.

The introduction of a nonzero $a$ modifies the equation of motion to
\beq
\ddot{\phi} \left( 1 + 24 a \frac{\dot{\phi}^2}{\phi^4} \right)- 24a \frac{\dot{\phi}^4}{\phi^5}+ 4 \lambda \phi^3 =0
\eeq
which has an important effect near the singularities.
Solving the equation taking $\phi(0) = \phi_0$ and $\dot{\phi}(0) = 0$, we find that instead of reaching the origin in finite time, the solution scales exponentially at early and late times:  $\phi( | t | \gg 1) \propto e^{-\left(\frac{1}{6 \lambda a}\right)^{1/4} |t|}$.
This is depicted by the blue curve in Fig.~\ref{fig:scalarcosmo}.

Again, it is crucial to note that there will generally be other higher-dimensional operators in the dilaton EFT, which will be important at large times as $\phi \rightarrow 0$. Still, it is interesting to note that a scalar analog of a puzzle in quantum gravity appears to be resolved by inclusion of the contribution from the conformal anomaly.

\section{Conclusions}
We have taken a holographic approach to studying the dilaton effective action and the $a$-anomaly in spontaneously broken CFTs. Central to our approach was studying the action in a derivative expansion, rather than an expansion in the fields. By integrating out KK gravitons in the RS model, we computed the radion effective action up to four derivatives. At the two-derivative level we obtained the radion kinetic and quartic terms. At the four-derivative level we identified the $(\partial \tau)^4$ term as well as novel couplings of the radion to light matter fields.

We identified the $a$-anomaly from the coefficient of the $(\partial \tau)^4$ term. As expected from anomaly matching arguments, our calculation of $a$ agrees with previous holographic computations of the $a$-anomaly in pure AdS without branes, in the large $N$ limit. We have also investigated the leading corrections to our result in the $1/N$ expansion, which correspond in the 5D picture to higher-curvature terms in the action. Again we found agreement with expectations from anomaly matching.

Another holographic setting in which one could explore the radion effective action is the ``soft wall'' background studied in Refs.~\cite{Csaki:2000wz,Bellazzini:2013fga}, involving a bulk scalar field with a flat potential. Like RS, this model spontaneously breaks scale invariance, so the low-energy effective action can likely be studied with the same techniques.

Notably, the dilaton effective action also includes $a$-dependent couplings of the dilaton to matter fields. These couplings are easier to measure at colliders than the four dilaton self-interaction but, like the $(\partial \tau)^4$ term, they too probe the $N$ of the CFT. In particular, we found a dimension-8 coupling of the form $T^{\mu\nu} \partial_\mu \tau \partial_\nu \tau$. The usual non-anomalous interactions of the dilaton only include couplings to the trace of the stress-energy tensor (like $\tau T$); the anomaly term induces a coupling to the traceless part of $T_{\mu\nu}$.

The anomaly term may also impact dilaton cosmology. We considered a toy model with a positive quartic dilaton potential, which pushes the dilaton field towards $\phi = 0$ (corresponding to $\tau \rightarrow \infty$, and a restoration of unbroken conformal symmetry). The $(\partial \tau)^4$ term induces an ``anomaly drag'' that prevents the dilaton field from rolling down to its minimum, fundamentally affecting the cosmology.

The dilaton effective theory provides an interesting scalar analog of gravity, as discussed in Ref.~\cite{Sundrum:2003yt}, admitting dilatonic black hole solutions and cosmologies. In the presence of the $a$-anomaly, we find the effective theory breaks down in the approach to the black hole horizon. This issue may or may not persist in the UV completion of the CFT; if it does, the $a$-anomaly is providing an indication that there are no time-independent dilatonic black holes. Dilaton cosmologies are modified by the $a$-term as well: without the $a$-anomaly, the equations of motion admit solutions where $\phi$ hits $0$ in finite time, but including the $a$-anomaly smooths out these singularities.

In this work, we have focused our attention on a very simple 5D model of spontaneous breaking of scale invariance. In particular we have not included any radius stabilization mechanism, and we have only considered matter fields localized purely on the IR brane. A more realistic model would include a stabilization mechanism and bulk fields, both of which would introduce explicit violation of conformal symmetry and deform the 5D geometry away from AdS. We expect that including these effects would not change the $(\partial \tau)^4$ term, since that is fixed by the $a$-anomaly, but the other terms in the effective action could be modified, and new explicit symmetry-breaking terms would arise.

Nevertheless, we have learned some general lessons from our results that should carry over, at the qualitative level, to more sophisticated models: 
\begin{itemize}
\item The dilaton effective action contains $a$-anomalous interactions that show up at four derivatives.
\item The $a$-anomalous interactions include the four-dilaton self-interaction as well as dilaton-matter couplings.
\item These interactions have implications for collider phenomenology and cosmology.
\end{itemize}

\begin{acknowledgments}
    We thank Brando Bellazzini for reading an early version of this draft, and providing useful comments. C.C., A.I. and F.S. are supported in part by the NSF grant PHY-2014071. C.C. is also supported in part by the BSF grant 2020220. J.H. is supported by the U.S. Department of Energy under Award Number DE-SC0009998. A.I. is supported by an NSERC PGS D fellowship (funding reference 557763). F.S. is also supported by a Klarman Fellowship at Cornell University.
    
 \end{acknowledgments}

\appendix

\section{RS radion action derivation}\label{sec:RScalculations}
Here we provide the details of the calculations leading to the RS radion effective action in \S~\ref{sec:holdilaton}.

\subsection{Order $\partial^2$}
The basic idea is to expand the 5D gravitational action using the metric ansatz in Eq.~\eqref{eq:metricansatzsimplified}:
\begin{equation}\label{eq:ansatzappendix}
    ds^2 = e^{-2 A} \eta_{\mu\nu} dx^\mu dx^\nu - \frac{A'^2}{k^2 - \frac{1}{2} e^{2 A}\left[\Box A - (\partial A)^2 \right]} dy^2 .
\end{equation}

We first expand the bulk Einstein--Hilbert action using this ansatz, assuming a cosmological constant $\Lambda = -6k^2$:
\begin{align}
    S_{\rm bulk} &= -\frac{1}{2\kappa^2} \int d^5 x \sqrt{g} \left( R + 2\Lambda \right) \nonumber \\
    &= \frac{1}{\kappa^2 k} \int d^5 x e^{-2A} \left[ A' \left(  \Box A- (\partial A)^2 \right) + 2 \partial A \cdot \partial A' - \Box A' \right] - 4k^2 A' e^{-4A} .
\end{align}
After integrating the boxes by parts, the Lagrangian density is seen to be a total $y$-derivative:
\beq
S_{\rm bulk} = \frac{1}{\kappa^2 k} \int d^5 x \partial_y \left[ k^2 e^{-4A} - \frac{1}{2} e^{-2A} (\partial A)^2 \right] .
\eeq

We can choose to either compactify the extra dimension on an interval $[y_0, y_1]$, or on a circle with the identification $y \Leftrightarrow -y$. (Recall that in our parametrization of the metric, the branes are held fixed at $y_0$ and $y_1$.) These approaches differ by a factor of $2$. Here we compactify on the interval and therefore we must include Gibbons--Hawking--York boundary terms:
\begin{equation}
    S_{\rm GHY} = -\frac{1}{\kappa^2} \int d^4 x \sqrt{h_0} K_0 - \frac{1}{\kappa^2} \int d^4 x \sqrt{h_1} K_1 ,
\end{equation}
where $h_i$ is the determinant of the induced metric on the brane at $y_i$, and $K_i$ is the scalar extrinsic curvature, related to the unit outward normal to the brane $\eta$ by $K = h^{AB} \nabla_A \eta_B$.  This gives
\beq
S_{\rm GHY} = \left. \frac{1}{\kappa^2k} \int d^4x \left[ -4 k^2e^{-4A} + 2 e^{-2A} (\partial A)^2 \right] \right|^{y_1}_{y_0} .
\eeq
If we chose instead to compactify on the full circle, these terms would appear in $S_{\rm bulk}$, arising from the discontinuity in $A'$ at $y_{0,1}$.

The off-diagonal $\mu$-$\nu$ Einstein equations are not yet satisfied at the linearized level.  To enforce these removes mixing with the tensor fluctuations $h_{\mu\nu}$, and isolates the scalar degree of freedom.  The off-diagonal $\mu$-$\nu$ equations yield
\beq
\partial_\mu \partial_\nu \partial_y e^{-2A} + \text{terms quadratic in fluctuations} = 0 .
\eeq
Diagonalization is thus achieved by taking $e^{-A_\text{UV}} = \sqrt{ e^{-2 A_\text{IR}} + e^{-2 A_0}-e^{-2 A_1}}$, where $A_{0,1} = \langle A(y_{0,1})\rangle$ are the background values of the warp factor on the branes.  More familiarly, in terms of a background $\langle A \rangle = ky$, and the linearized radion fluctuation, $F(x,y) = A(x,y)-k y \approx f(y) r(x) $, the $\mu$-$\nu$ equation above gives the usual equation for the radion wave-function, $\partial_y \left( e^{-2ky} f(y) \right) = 0$.

Lastly, we add brane tensions $\lambda_0$ and $\lambda_1$, leading to the brane-localized action
\begin{equation}\begin{split}
    S_{\rm brane} &= - \int d^4 x \sqrt{h_0} \lambda_0 - \int d^4 x \sqrt{h_1} \lambda_1 \\
    &= - \int d^4 x \left[ \lambda_0 e^{-4A_{\rm UV}} +  \lambda_1 e^{-4A_{\rm IR}} \right] .
\end{split}\end{equation}

The effective action at two derivatives is then, as quoted in Eq.~\eqref{eq:effaction},
\begin{equation}\begin{split}
    S_\text{radion} &= 2 S_{\rm bulk} + 2 S_{\rm GHY} + S_{\rm brane} \\
    &= \frac{3}{\kappa^2 k} \int d^4x\left. e^{-2A} (\partial A)^2  \right|^{y_1}_{y_0} -\int d^4x \left[ \lambda_1 + \frac{6 k}{\kappa^2} \right]  e^{-4A_{\rm IR}} - \left[ \lambda_0 - \frac{6 k}{\kappa^2} \right] e^{-4A_{\rm UV}} .
\end{split}\end{equation}
Note the factors of two which ensure agreement with the theory compactified on the full circle.

\subsection{Order $\partial^4$}
At the next order in derivatives we must include the tensor fluctuation in the metric ansatz, that is,
\begin{equation}
    ds^2 = e^{-2 A(x,y)} \left(\eta_{\mu\nu}+h_{\mu\nu}(x,y)\right) dx^\mu dx^\nu - B^2(x,y) dy^2 .
\end{equation}
We now proceed to solve the classical equation of motion for $h_{\mu\nu}$ up to second order in derivatives, which corresponds to linear order in $h_{\mu\nu}$.

Recall the $y$-$y$ component of the Einstein equations relates $A$ and $B$ by $(A'/B)^2 = k^2 - \frac{1}{2} e^{2A} \left[ \Box A - (\partial A)^2 \right]$.
The $\mu$-$\nu$ Einstein equations including the tensor at leading order are a total $y$-derivative. This leads to an equation of motion for the tensor,
\begin{equation}\label{eq:heom}
\partial_y \left( \frac{e^{-4A}}{A'} h'_{\mu\nu} \right) =\frac{2}{k^2} \partial_y \left[ e^{-2A} \left( \partial_\mu \partial_\nu A + \partial_\mu A \partial_\nu A + \frac{1}{2} \eta_{\mu\nu} \left((\partial A)^2 - 2 \Box A \right) \right) \right] \equiv 2 \partial_y J_{\mu\nu} .
\end{equation}
This equation is exact at second order in the derivative expansion.

The boundary condition on $h_{\mu\nu}$ is determined by the requirement that the variation of the boundary terms in the action (namely, the Gibbons--Hawking--York term and the matter action) vanishes. We find
\begin{equation}\label{eq:hbc}
    \frac{e^{-4A}}{A'} h_{\mu\nu}' \Big |_{y_1} = \frac{3}{2k^2} e^{-2A_{\rm IR}} \left( \Box A_{\rm IR} - (\partial A_{\rm IR})^2 \right) \eta_{\mu\nu} - \frac{\kappa^2}{k} T_{\mu\nu} .
\end{equation}
Taking the trace of the boundary condition and using the fact that $h_{\mu\nu}$ is traceless yields the radion equation of motion at two derivatives:
\begin{equation}
    e^{-2A_{\rm IR}} \left( \Box A_{\rm IR} - (\partial A_{\rm IR})^2 \right) = \frac{\kappa^2 k}{6} T .
\end{equation}

To determine the effective action after integrating out the tensor, which is equivalent to summing over the KK graviton excitations, we expand the action to second order in $h_{\mu\nu}$.  As we are concerned only with terms with four derivatives or less, we do not consider terms containing 4D derivatives of $h$.  We add together the contributions from the bulk Einstein--Hilbert action, the Gibbons--Hawking--York boundary terms, the brane tension, and the matter action. This final action is given by:
\begin{equation}\begin{split}
    S_\text{tensor} &=-\frac{k}{4 \kappa^2}\int d^5 x \left[\frac{e^{-4A}}{A'} (h'_{\mu\nu})^2 + 4 h^{\mu\nu} \partial_y J_{\mu\nu} \right] \\
    &+ \frac{1}{4} \left( \frac{6k}{\kappa^2}+ \lambda_1 \right) \int d^4 x e^{-4A_{\rm IR}} (h_{\mu\nu})^2 - \frac{1}{2} \int d^4 x h_{\mu\nu} \left( T^{\mu\nu} - \frac{1}{4} \eta^{\mu\nu} T \right)
\end{split}\end{equation}
where indices are raised and lowered with the Minkowski metric.

It is easy to solve for $h'$ using Eq.~\eqref{eq:heom} and Eq.~\eqref{eq:hbc}:
\begin{equation}\label{eq:RSgravsolution}
    h'_{\mu\nu} =  A' e^{4 A} \left[ 2 J_{\mu\nu}(y_1) - 2 J_{\mu\nu}(y) - \frac{\kappa^2}{k} \left( T_{\mu\nu} - \frac{1}{4} \eta_{\mu\nu} T \right) \right] .
\end{equation}
Substituting this solution back into the action and integrating by parts then leads to
\begin{equation}\label{eq:RSgravaction}
    S_\text{tensor} = \frac{k}{4\kappa^2} \int d^5 x e^{4 A} A' \left[ 2 J_{\mu\nu}(y_1) - 2 J_{\mu\nu}(y) - \frac{\kappa^2}{k} \left( T_{\mu\nu} - \frac{1}{4} \eta_{\mu\nu} T \right) \right]^2 .
\end{equation}
Expanding this in terms of $A$, integrating by parts with respect to the 4D derivatives, and employing the two-derivative radion equation of motion, this expression reduces to a pure boundary term:
\begin{equation}\begin{split}
    S_\text{tensor} =  \frac{1}{4 \kappa^2k^3} &\int d^5 x \partial_y \left[ (\partial A)^4 + \kappa^2 k e^{2A} \partial_\mu A \partial_\nu A \left(T^{\mu\nu} - \frac{1}{6} \eta^{\mu\nu} T \right) \right] \\
    + \frac{\kappa^2}{16k} &\int d^4 x  \left( T_{\mu\nu} - \frac{1}{4} \eta_{\mu\nu} T \right)^2 .
\end{split}\end{equation}
Upon performing the integral over $y$, we obtain Eq.~\eqref{eq:RStensoraction}.

\section{Subleading bulk contributions to the holographic action}\label{sec:GBgravity}
Here we provide the details of including higher-derivative terms in the bulk action and the resulting effects on the $a$-anomaly. The most general four-derivative bulk Lagrangian is
\begin{equation}
    c_1 R^2 + c_2 R_{ab}^2 + c_3 R_{abcd}^2,
\end{equation}
where we are ignoring a possible $\nabla^2 R$ term because it is a total derivative. One can perform a field redefinition of the metric to arbitrarily shift $c_1$ and $c_2$ (but not $c_3$)~\cite{Cheung:2018cwt}. In this way, the Lagrangian can be put into the form
\begin{equation}\label{eq:GBfieldredef}
    c_3 \left( R^2 - 4 R_{ab}^2 + R_{abcd}^2 \right) .
\end{equation}
The quantity in brackets is a Gauss--Bonnet term, the generalization of the 4D Euler density. This parametrization of the four-derivative terms is especially convenient for our purposes.

The field redefinition needed to put the four-derivative Lagrangian in the form of Eq.~\eqref{eq:GBfieldredef} necessarily introduces new couplings of the form $T_{ab}^2$. However, since we are concerned with a bulk theory with no matter except for a cosmological constant, we have $T_{ab} \propto g_{ab}$. Hence $T_{ab}^2$ is just a constant, which may be absorbed into a shift of the cosmological constant.

Motivated by the discussion above, we add to the bulk action a Gauss--Bonnet term
\begin{equation}\label{eq:GBbulk}
    S_{\rm GB,bulk} = \frac{\lambda_{\rm GB}}{2\kappa^2 k^2} \left[ \frac{2\kappa^2 k^3}{24\pi^3} \right]^{2/3} \int d^5 x \sqrt{g} \left( R^2 - 4 R_{ab}^2 + R_{abcd}^2 \right)
\end{equation}
with $\lambda_{\rm GB}$ is a dimensionless constant. Recall the overall coefficient is expected from na\"ive dimensional analysis.
Expanding the bulk Gauss--Bonnet term to order $\partial^4$ using the metric ansatz Eq.~\eqref{eq:ansatzappendix}, we find
\begin{equation}\begin{split}
    S_{\rm GB,bulk} = \frac{4\lambda_{\rm GB}}{\kappa^2 k^3} \left[ \frac{2\kappa^2 k^3}{24\pi^3} \right]^{2/3} \int d^5 x & \: 15 k^4 A' e^{-4A} +  \frac{15}{4} e^{-2A} k^2 A' \left( \Box A - (\partial A)^2 \right) \\
    &+A' \left[ -2 \partial^\mu A \partial^\nu A \partial_\mu \partial_\nu A - (\partial A)^2 \Box A + (\Box A)^2 - (\partial_\mu \partial_\nu A)^2 \right] \\
    &+ (\partial A)^2 \Box A' + 2 \partial A \cdot \partial A' \Box A - 2 \Box A \Box A' + 4 \partial^\mu A' \partial^\nu A \partial_\mu \partial_\nu A \\
    &+ 2 \partial^\mu A \partial^\nu A \partial_\mu \partial_\nu A' + 2 \partial_\mu \partial_\nu A \partial^\mu \partial^\nu A' .
\end{split}\end{equation}
This turns out to be a total $y$-derivative after integrating the boxes by parts. Performing the integral over $y$ and assuming the UV brane has been sent to the AdS boundary, we find
\begin{equation}\begin{split}
    S_{\rm GB,bulk} = \frac{\lambda_{\rm GB}}{\kappa^2 k^3} \left[ \frac{2\kappa^2 k^3}{24\pi^3} \right]^{2/3} &\int d^4 x  \left[ 15 k^4 e^{-4A_{\rm IR}} - \frac{15}{2} k^2 e^{-2A_{\rm IR}} (\partial A_{\rm IR})^2 + (\partial A_{\rm IR})^4 \right. \\
    &+ \left. 4 \partial^\mu A_{\rm IR} \partial^\nu A_{\rm IR} \partial_\mu \partial_\nu A_{\rm IR} \right] .
\end{split}\end{equation}
One can use integration by parts and the equation of motion to show the last term can be replaced by $-2(\partial A_{\rm IR})^4$, at the level of four derivatives.

We must also include the appropriate Gibbons--Hawking--York boundary term~\cite{Liu:2008zf}:
\begin{equation}
    S_{\rm GB,GHY} = -\frac{2\lambda_{\rm GB}}{\kappa^2 k^2} \left[ \frac{2\kappa^2 k^3}{24\pi^3} \right]^{2/3} \int d^4 x \sqrt{h_1} \left[ 2 \mathcal{G}_{AB} K^{AB} + \frac{1}{3} \left( K^3 - 3 K K_{AB}^2 + 2 K_A^B K_B^C K_C^A \right) \right] 
\end{equation}
where $\mathcal{G}_{AB}$ is the Einstein tensor computed from the induced metric on the boundary. After an integration by parts, one obtains
\begin{equation}
    S_{\rm GB,GHY} = \frac{\lambda_{\rm GB}}{\kappa^2 k^3} \left[ \frac{2\kappa^2 k^3}{24\pi^3} \right]^{2/3} \int d^4 x \left[ -16 k^4 e^{-4A_{\rm IR}} + 36k^2 e^{-2A_{\rm IR}} (\partial A_{\rm IR})^2 \right] .
\end{equation}
Hence, the Gauss--Bonnet term contributes to the effective radion action as
\begin{equation}\label{eq:GBtermeffaction}\begin{split}
    2S_{\rm GB,bulk} &+ 2S_{\rm GB,GHY} \\
    &= \frac{\lambda_{\rm GB}}{\kappa^2 k^3} \left[ \frac{2\kappa^2 k^3}{24\pi^3} \right]^{2/3} \int d^4 x \left[ -2 k^4 e^{-4A_{\rm IR}} + 57k^2 e^{-2A_{\rm IR}} (\partial A_{\rm IR})^2 - 2(\partial A_{\rm IR})^4 \right] .
\end{split}\end{equation}
Recall the factors of two ensure agreement with the theory compactified on the full circle.

There are three ramifications of the Gauss--Bonnet term, exemplified by Eq.~\eqref{eq:GBtermeffaction}. First, there is a contribution to the radion quartic. One can restore the tuning of the quartic by shifting the IR brane tension as $\tau_{\rm IR} \rightarrow \tau_{\rm IR} + \Delta \tau_{\rm IR}$, where
\begin{equation}
    \Delta \tau_{\rm IR} = -\frac{2\lambda_{\rm GB} k}{\kappa^2} \left[ \frac{2\kappa^2 k^3}{24\pi^3} \right]^{2/3} .
\end{equation}
Second, the Gauss--Bonnet term modifies the dilaton decay constant,
\begin{equation}
    \Delta f^2 = e^{-2A_1} \frac{57 \lambda_{\rm GB}}{\kappa^2 k} \left[ \frac{2\kappa^2 k^3}{24\pi^3} \right]^{2/3} .
\end{equation}
Most importantly, we see from Eq.~\eqref{eq:GBtermeffaction} that there is a direct contribution to the $a$-anomaly:
\begin{equation}\label{eq:GBanomalydirect}
    \Delta a = -\frac{\lambda_{\rm GB}}{\kappa^2 k^3} \left[ \frac{2\kappa^2 k^3}{24\pi^3} \right]^{2/3}.
\end{equation}
However, the new $a$-anomaly is not simply Eq.~\eqref{eq:RSanomaly} plus $\Delta a$. This is because the effective quadratic action for the graviton is also modified by the Gauss--Bonnet term, which alters the contribution to the $a$-anomaly obtained from integrating out KK gravitons.

To understand this further, we write the field equations (including the Gauss--Bonnet term) as $0 = G_{mn} - \kappa^2 T_{mn} - \frac{\lambda_{\rm GB}}{k^2} \left[ \frac{2\kappa^2 k^3}{24\pi^3} \right]^{2/3}I_{mn}$, where
\begin{equation}
    I_{mn} =  -\frac{1}{2} \left( R^2 - 4R_{ab}^2 + R_{abcd}^2 \right)g_{mn} - 4 R_m^a R_{na} + 2 R R_{mn} - 4 R^{ab} R_{manb} + 2 R_m^{\: abc} R_{nabc} .
\end{equation}
It is easy to see that these field equations admit AdS vacua (corresponding to e.g. $\langle A \rangle = k y$), with a bulk cosmological constant
\begin{equation}
    \Lambda = -6k^2 \left(1 - 2 \lambda_{\rm GB} \left[ \frac{2\kappa^2 k^3}{24\pi^3} \right]^{2/3}\right) .
\end{equation}
The presence of the Gauss--Bonnet term shifts the cosmological constant. As in the Einstein gravity case, one solves for the radion wavefunction and then the KK gravitons. We find that the radion wavefunction is unchanged, that is, $A(x,y) \approx k y + e^{2ky} r(x)$. The graviton equation of motion is also unchanged from Eq.~\eqref{eq:heom}, up to source terms with four derivatives, which cannot affect the $a$-anomaly term. Hence, the solution for $h_{\mu\nu}$ (Eq.~\eqref{eq:RSgravsolution}) is unchanged by the presence of the Gauss--Bonnet term. After expanding the action to quadratic order in the graviton, one finds that it is of the same form as Eq.~\eqref{eq:RSgravaction}, but with an additional factor of $1 - 4 \lambda_{\rm GB}\left[ \frac{2\kappa^2 k^3}{24\pi^3} \right]^{2/3}$.
Then the contribution to the $a$-anomaly from integrating out KK gravitons is
\begin{equation}
    a_{\rm KK} = \frac{1}{8\kappa^2 k^3}\left( 1 - 4\lambda_{\rm GB} \left[ \frac{2\kappa^2 k^3}{24\pi^3} \right]^{2/3} \right).
\end{equation}
Adding this to the direct contribution in Eq.~\eqref{eq:GBanomalydirect}, we find the $a$-anomaly with the Gauss--Bonnet term included is
\begin{equation}
    a_{\rm GB} = a_{\rm RS} \left( 1 - 12 \lambda_{\rm GB}\left[ \frac{2\kappa^2 k^3}{24\pi^3} \right]^{2/3} \right)
\end{equation}
where $a_{\rm RS} = 1/(8\kappa^2k^3)$ is the $a$-anomaly without the Gauss--Bonnet term.

\bibliography{References}

\begin{thebibliography}{54}%
\makeatletter
\providecommand \@ifxundefined [1]{%
 \@ifx{#1\undefined}
}%
\providecommand \@ifnum [1]{%
 \ifnum #1\expandafter \@firstoftwo
 \else \expandafter \@secondoftwo
 \fi
}%
\providecommand \@ifx [1]{%
 \ifx #1\expandafter \@firstoftwo
 \else \expandafter \@secondoftwo
 \fi
}%
\providecommand \natexlab [1]{#1}%
\providecommand \enquote  [1]{``#1''}%
\providecommand \bibnamefont  [1]{#1}%
\providecommand \bibfnamefont [1]{#1}%
\providecommand \citenamefont [1]{#1}%
\providecommand \href@noop [0]{\@secondoftwo}%
\providecommand \href [0]{\begingroup \@sanitize@url \@href}%
\providecommand \@href[1]{\@@startlink{#1}\@@href}%
\providecommand \@@href[1]{\endgroup#1\@@endlink}%
\providecommand \@sanitize@url [0]{\catcode `\\12\catcode `\$12\catcode
  `\&12\catcode `\#12\catcode `\^12\catcode `\_12\catcode `\%12\relax}%
\providecommand \@@startlink[1]{}%
\providecommand \@@endlink[0]{}%
\providecommand \url  [0]{\begingroup\@sanitize@url \@url }%
\providecommand \@url [1]{\endgroup\@href {#1}{\urlprefix }}%
\providecommand \urlprefix  [0]{URL }%
\providecommand \Eprint [0]{\href }%
\providecommand \doibase [0]{http://dx.doi.org/}%
\providecommand \selectlanguage [0]{\@gobble}%
\providecommand \bibinfo  [0]{\@secondoftwo}%
\providecommand \bibfield  [0]{\@secondoftwo}%
\providecommand \translation [1]{[#1]}%
\providecommand \BibitemOpen [0]{}%
\providecommand \bibitemStop [0]{}%
\providecommand \bibitemNoStop [0]{.\EOS\space}%
\providecommand \EOS [0]{\spacefactor3000\relax}%
\providecommand \BibitemShut  [1]{\csname bibitem#1\endcsname}%
\let\auto@bib@innerbib\@empty
\bibitem [{\citenamefont {Agashe}\ \emph {et~al.}(2005)\citenamefont {Agashe},
  \citenamefont {Contino},\ and\ \citenamefont {Pomarol}}]{Agashe:2004rs}%
  \BibitemOpen
  \bibfield  {author} {\bibinfo {author} {\bibfnamefont {K.}~\bibnamefont
  {Agashe}}, \bibinfo {author} {\bibfnamefont {R.}~\bibnamefont {Contino}}, \
  and\ \bibinfo {author} {\bibfnamefont {A.}~\bibnamefont {Pomarol}},\ }\href
  {\doibase 10.1016/j.nuclphysb.2005.04.035} {\bibfield  {journal} {\bibinfo
  {journal} {Nucl. Phys. B}\ }\textbf {\bibinfo {volume} {719}},\ \bibinfo
  {pages} {165} (\bibinfo {year} {2005})},\ \Eprint
  {http://arxiv.org/abs/hep-ph/0412089} {arXiv:hep-ph/0412089} \BibitemShut
  {NoStop}%
\bibitem [{\citenamefont {Contino}(2011)}]{Contino:2010rs}%
  \BibitemOpen
  \bibfield  {author} {\bibinfo {author} {\bibfnamefont {R.}~\bibnamefont
  {Contino}},\ }in\ \href {\doibase 10.1142/9789814327183_0005} {\emph
  {\bibinfo {booktitle} {{Theoretical Advanced Study Institute in Elementary
  Particle Physics}: {Physics of the Large and the Small}}}}\ (\bibinfo {year}
  {2011})\ pp.\ \bibinfo {pages} {235--306},\ \Eprint
  {http://arxiv.org/abs/1005.4269} {arXiv:1005.4269 [hep-ph]} \BibitemShut
  {NoStop}%
\bibitem [{\citenamefont {Bellazzini}\ \emph
  {et~al.}(2014{\natexlab{a}})\citenamefont {Bellazzini}, \citenamefont
  {Cs\'aki},\ and\ \citenamefont {Serra}}]{Bellazzini:2014yua}%
  \BibitemOpen
  \bibfield  {author} {\bibinfo {author} {\bibfnamefont {B.}~\bibnamefont
  {Bellazzini}}, \bibinfo {author} {\bibfnamefont {C.}~\bibnamefont {Cs\'aki}},
  \ and\ \bibinfo {author} {\bibfnamefont {J.}~\bibnamefont {Serra}},\ }\href
  {\doibase 10.1140/epjc/s10052-014-2766-x} {\bibfield  {journal} {\bibinfo
  {journal} {Eur. Phys. J. C}\ }\textbf {\bibinfo {volume} {74}},\ \bibinfo
  {pages} {2766} (\bibinfo {year} {2014}{\natexlab{a}})},\ \Eprint
  {http://arxiv.org/abs/1401.2457} {arXiv:1401.2457 [hep-ph]} \BibitemShut
  {NoStop}%
\bibitem [{\citenamefont {Panico}\ and\ \citenamefont
  {Wulzer}(2016)}]{Panico:2015jxa}%
  \BibitemOpen
  \bibfield  {author} {\bibinfo {author} {\bibfnamefont {G.}~\bibnamefont
  {Panico}}\ and\ \bibinfo {author} {\bibfnamefont {A.}~\bibnamefont
  {Wulzer}},\ }\href {\doibase 10.1007/978-3-319-22617-0} {\emph {\bibinfo
  {title} {{The Composite Nambu-Goldstone Higgs}}}},\ Vol.\ \bibinfo {volume}
  {913}\ (\bibinfo  {publisher} {Springer},\ \bibinfo {year} {2016})\ \Eprint
  {http://arxiv.org/abs/1506.01961} {arXiv:1506.01961 [hep-ph]} \BibitemShut
  {NoStop}%
\bibitem [{\citenamefont {Gherghetta}\ and\ \citenamefont {von
  Harling}(2010)}]{Gherghetta:2010cq}%
  \BibitemOpen
  \bibfield  {author} {\bibinfo {author} {\bibfnamefont {T.}~\bibnamefont
  {Gherghetta}}\ and\ \bibinfo {author} {\bibfnamefont {B.}~\bibnamefont {von
  Harling}},\ }\href {\doibase 10.1007/JHEP04(2010)039} {\bibfield  {journal}
  {\bibinfo  {journal} {JHEP}\ }\textbf {\bibinfo {volume} {04}},\ \bibinfo
  {pages} {039} (\bibinfo {year} {2010})},\ \Eprint
  {http://arxiv.org/abs/1002.2967} {arXiv:1002.2967 [hep-ph]} \BibitemShut
  {NoStop}%
\bibitem [{\citenamefont {von Harling}\ and\ \citenamefont
  {McDonald}(2012)}]{vonHarling:2012sz}%
  \BibitemOpen
  \bibfield  {author} {\bibinfo {author} {\bibfnamefont {B.}~\bibnamefont {von
  Harling}}\ and\ \bibinfo {author} {\bibfnamefont {K.~L.}\ \bibnamefont
  {McDonald}},\ }\href {\doibase 10.1007/JHEP08(2012)048} {\bibfield  {journal}
  {\bibinfo  {journal} {JHEP}\ }\textbf {\bibinfo {volume} {08}},\ \bibinfo
  {pages} {048} (\bibinfo {year} {2012})},\ \Eprint
  {http://arxiv.org/abs/1203.6646} {arXiv:1203.6646 [hep-ph]} \BibitemShut
  {NoStop}%
\bibitem [{\citenamefont {Hong}\ \emph {et~al.}(2020)\citenamefont {Hong},
  \citenamefont {Kurup},\ and\ \citenamefont {Perelstein}}]{Hong:2019nwd}%
  \BibitemOpen
  \bibfield  {author} {\bibinfo {author} {\bibfnamefont {S.}~\bibnamefont
  {Hong}}, \bibinfo {author} {\bibfnamefont {G.}~\bibnamefont {Kurup}}, \ and\
  \bibinfo {author} {\bibfnamefont {M.}~\bibnamefont {Perelstein}},\ }\href
  {\doibase 10.1103/PhysRevD.101.095037} {\bibfield  {journal} {\bibinfo
  {journal} {Phys. Rev. D}\ }\textbf {\bibinfo {volume} {101}},\ \bibinfo
  {pages} {095037} (\bibinfo {year} {2020})},\ \Eprint
  {http://arxiv.org/abs/1910.10160} {arXiv:1910.10160 [hep-ph]} \BibitemShut
  {NoStop}%
\bibitem [{\citenamefont {Chaffey}\ \emph {et~al.}(2021)\citenamefont
  {Chaffey}, \citenamefont {Fichet},\ and\ \citenamefont
  {Tanedo}}]{Chaffey:2021tmj}%
  \BibitemOpen
  \bibfield  {author} {\bibinfo {author} {\bibfnamefont {I.}~\bibnamefont
  {Chaffey}}, \bibinfo {author} {\bibfnamefont {S.}~\bibnamefont {Fichet}}, \
  and\ \bibinfo {author} {\bibfnamefont {P.}~\bibnamefont {Tanedo}},\ }\href
  {\doibase 10.1007/JHEP06(2021)008} {\bibfield  {journal} {\bibinfo  {journal}
  {JHEP}\ }\textbf {\bibinfo {volume} {06}},\ \bibinfo {pages} {008} (\bibinfo
  {year} {2021})},\ \Eprint {http://arxiv.org/abs/2102.05674} {arXiv:2102.05674
  [hep-ph]} \BibitemShut {NoStop}%
\bibitem [{\citenamefont {Cs\'aki}\ \emph
  {et~al.}(2021{\natexlab{a}})\citenamefont {Cs\'aki}, \citenamefont {Hong},
  \citenamefont {Kurup}, \citenamefont {Lee}, \citenamefont {Perelstein},\ and\
  \citenamefont {Xue}}]{Csaki:2021gfm}%
  \BibitemOpen
  \bibfield  {author} {\bibinfo {author} {\bibfnamefont {C.}~\bibnamefont
  {Cs\'aki}}, \bibinfo {author} {\bibfnamefont {S.}~\bibnamefont {Hong}},
  \bibinfo {author} {\bibfnamefont {G.}~\bibnamefont {Kurup}}, \bibinfo
  {author} {\bibfnamefont {S.~J.}\ \bibnamefont {Lee}}, \bibinfo {author}
  {\bibfnamefont {M.}~\bibnamefont {Perelstein}}, \ and\ \bibinfo {author}
  {\bibfnamefont {W.}~\bibnamefont {Xue}},\ }\href@noop {} {\  (\bibinfo {year}
  {2021}{\natexlab{a}})},\ \Eprint {http://arxiv.org/abs/2105.07035}
  {arXiv:2105.07035 [hep-ph]} \BibitemShut {NoStop}%
\bibitem [{\citenamefont {Cs\'aki}\ \emph
  {et~al.}(2021{\natexlab{b}})\citenamefont {Cs\'aki}, \citenamefont {Hong},
  \citenamefont {Kurup}, \citenamefont {Lee}, \citenamefont {Perelstein},\ and\
  \citenamefont {Xue}}]{Csaki:2021xpy}%
  \BibitemOpen
  \bibfield  {author} {\bibinfo {author} {\bibfnamefont {C.}~\bibnamefont
  {Cs\'aki}}, \bibinfo {author} {\bibfnamefont {S.}~\bibnamefont {Hong}},
  \bibinfo {author} {\bibfnamefont {G.}~\bibnamefont {Kurup}}, \bibinfo
  {author} {\bibfnamefont {S.~J.}\ \bibnamefont {Lee}}, \bibinfo {author}
  {\bibfnamefont {M.}~\bibnamefont {Perelstein}}, \ and\ \bibinfo {author}
  {\bibfnamefont {W.}~\bibnamefont {Xue}},\ }\href@noop {} {\  (\bibinfo {year}
  {2021}{\natexlab{b}})},\ \Eprint {http://arxiv.org/abs/2105.14023}
  {arXiv:2105.14023 [hep-ph]} \BibitemShut {NoStop}%
\bibitem [{\citenamefont {Cs\'aki}\ \emph {et~al.}(2019)\citenamefont
  {Cs\'aki}, \citenamefont {Lee}, \citenamefont {Lee}, \citenamefont
  {Lombardo},\ and\ \citenamefont {Telem}}]{Csaki:2018kxb}%
  \BibitemOpen
  \bibfield  {author} {\bibinfo {author} {\bibfnamefont {C.}~\bibnamefont
  {Cs\'aki}}, \bibinfo {author} {\bibfnamefont {G.}~\bibnamefont {Lee}},
  \bibinfo {author} {\bibfnamefont {S.~J.}\ \bibnamefont {Lee}}, \bibinfo
  {author} {\bibfnamefont {S.}~\bibnamefont {Lombardo}}, \ and\ \bibinfo
  {author} {\bibfnamefont {O.}~\bibnamefont {Telem}},\ }\href {\doibase
  10.1007/JHEP03(2019)142} {\bibfield  {journal} {\bibinfo  {journal} {JHEP}\
  }\textbf {\bibinfo {volume} {03}},\ \bibinfo {pages} {142} (\bibinfo {year}
  {2019})},\ \Eprint {http://arxiv.org/abs/1811.06019} {arXiv:1811.06019
  [hep-ph]} \BibitemShut {NoStop}%
\bibitem [{\citenamefont {Bloch}\ \emph {et~al.}(2020)\citenamefont {Bloch},
  \citenamefont {Cs\'aki}, \citenamefont {Geller},\ and\ \citenamefont
  {Volansky}}]{Bloch:2019bvc}%
  \BibitemOpen
  \bibfield  {author} {\bibinfo {author} {\bibfnamefont {I.~M.}\ \bibnamefont
  {Bloch}}, \bibinfo {author} {\bibfnamefont {C.}~\bibnamefont {Cs\'aki}},
  \bibinfo {author} {\bibfnamefont {M.}~\bibnamefont {Geller}}, \ and\ \bibinfo
  {author} {\bibfnamefont {T.}~\bibnamefont {Volansky}},\ }\href {\doibase
  10.1007/JHEP12(2020)191} {\bibfield  {journal} {\bibinfo  {journal} {JHEP}\
  }\textbf {\bibinfo {volume} {12}},\ \bibinfo {pages} {191} (\bibinfo {year}
  {2020})},\ \Eprint {http://arxiv.org/abs/1912.08840} {arXiv:1912.08840
  [hep-ph]} \BibitemShut {NoStop}%
\bibitem [{\citenamefont {Cs\'aki}\ \emph
  {et~al.}(2021{\natexlab{c}})\citenamefont {Cs\'aki}, \citenamefont
  {D'Agnolo}, \citenamefont {Geller},\ and\ \citenamefont
  {Ismail}}]{Csaki:2020zqz}%
  \BibitemOpen
  \bibfield  {author} {\bibinfo {author} {\bibfnamefont {C.}~\bibnamefont
  {Cs\'aki}}, \bibinfo {author} {\bibfnamefont {R.~T.}\ \bibnamefont
  {D'Agnolo}}, \bibinfo {author} {\bibfnamefont {M.}~\bibnamefont {Geller}}, \
  and\ \bibinfo {author} {\bibfnamefont {A.}~\bibnamefont {Ismail}},\ }\href
  {\doibase 10.1103/PhysRevLett.126.091801} {\bibfield  {journal} {\bibinfo
  {journal} {Phys. Rev. Lett.}\ }\textbf {\bibinfo {volume} {126}},\ \bibinfo
  {pages} {091801} (\bibinfo {year} {2021}{\natexlab{c}})},\ \Eprint
  {http://arxiv.org/abs/2007.14396} {arXiv:2007.14396 [hep-ph]} \BibitemShut
  {NoStop}%
\bibitem [{\citenamefont {Dymarsky}\ \emph {et~al.}(2015)\citenamefont
  {Dymarsky}, \citenamefont {Komargodski}, \citenamefont {Schwimmer},\ and\
  \citenamefont {Theisen}}]{Dymarsky:2013pqa}%
  \BibitemOpen
  \bibfield  {author} {\bibinfo {author} {\bibfnamefont {A.}~\bibnamefont
  {Dymarsky}}, \bibinfo {author} {\bibfnamefont {Z.}~\bibnamefont
  {Komargodski}}, \bibinfo {author} {\bibfnamefont {A.}~\bibnamefont
  {Schwimmer}}, \ and\ \bibinfo {author} {\bibfnamefont {S.}~\bibnamefont
  {Theisen}},\ }\href {\doibase 10.1007/JHEP10(2015)171} {\bibfield  {journal}
  {\bibinfo  {journal} {JHEP}\ }\textbf {\bibinfo {volume} {10}},\ \bibinfo
  {pages} {171} (\bibinfo {year} {2015})},\ \Eprint
  {http://arxiv.org/abs/1309.2921} {arXiv:1309.2921 [hep-th]} \BibitemShut
  {NoStop}%
\bibitem [{\citenamefont {Nakayama}(2015)}]{Nakayama:2013is}%
  \BibitemOpen
  \bibfield  {author} {\bibinfo {author} {\bibfnamefont {Y.}~\bibnamefont
  {Nakayama}},\ }\href {\doibase 10.1016/j.physrep.2014.12.003} {\bibfield
  {journal} {\bibinfo  {journal} {Phys. Rept.}\ }\textbf {\bibinfo {volume}
  {569}},\ \bibinfo {pages} {1} (\bibinfo {year} {2015})},\ \Eprint
  {http://arxiv.org/abs/1302.0884} {arXiv:1302.0884 [hep-th]} \BibitemShut
  {NoStop}%
\bibitem [{\citenamefont {Randall}\ and\ \citenamefont
  {Sundrum}(1999)}]{Randall:1999ee}%
  \BibitemOpen
  \bibfield  {author} {\bibinfo {author} {\bibfnamefont {L.}~\bibnamefont
  {Randall}}\ and\ \bibinfo {author} {\bibfnamefont {R.}~\bibnamefont
  {Sundrum}},\ }\href {\doibase 10.1103/PhysRevLett.83.3370} {\bibfield
  {journal} {\bibinfo  {journal} {Phys. Rev. Lett.}\ }\textbf {\bibinfo
  {volume} {83}},\ \bibinfo {pages} {3370} (\bibinfo {year} {1999})},\ \Eprint
  {http://arxiv.org/abs/hep-ph/9905221} {arXiv:hep-ph/9905221} \BibitemShut
  {NoStop}%
\bibitem [{\citenamefont {Goldberger}\ and\ \citenamefont
  {Wise}(1999)}]{Goldberger:1999uk}%
  \BibitemOpen
  \bibfield  {author} {\bibinfo {author} {\bibfnamefont {W.~D.}\ \bibnamefont
  {Goldberger}}\ and\ \bibinfo {author} {\bibfnamefont {M.~B.}\ \bibnamefont
  {Wise}},\ }\href {\doibase 10.1103/PhysRevLett.83.4922} {\bibfield  {journal}
  {\bibinfo  {journal} {Phys. Rev. Lett.}\ }\textbf {\bibinfo {volume} {83}},\
  \bibinfo {pages} {4922} (\bibinfo {year} {1999})},\ \Eprint
  {http://arxiv.org/abs/hep-ph/9907447} {arXiv:hep-ph/9907447} \BibitemShut
  {NoStop}%
\bibitem [{\citenamefont {Csaki}\ \emph
  {et~al.}(2000{\natexlab{a}})\citenamefont {Csaki}, \citenamefont {Graesser},
  \citenamefont {Randall},\ and\ \citenamefont {Terning}}]{Csaki:1999mp}%
  \BibitemOpen
  \bibfield  {author} {\bibinfo {author} {\bibfnamefont {C.}~\bibnamefont
  {Csaki}}, \bibinfo {author} {\bibfnamefont {M.}~\bibnamefont {Graesser}},
  \bibinfo {author} {\bibfnamefont {L.}~\bibnamefont {Randall}}, \ and\
  \bibinfo {author} {\bibfnamefont {J.}~\bibnamefont {Terning}},\ }\href
  {\doibase 10.1103/PhysRevD.62.045015} {\bibfield  {journal} {\bibinfo
  {journal} {Phys. Rev. D}\ }\textbf {\bibinfo {volume} {62}},\ \bibinfo
  {pages} {045015} (\bibinfo {year} {2000}{\natexlab{a}})},\ \Eprint
  {http://arxiv.org/abs/hep-ph/9911406} {arXiv:hep-ph/9911406} \BibitemShut
  {NoStop}%
\bibitem [{\citenamefont {Goldberger}\ and\ \citenamefont
  {Wise}(2000)}]{Goldberger:1999un}%
  \BibitemOpen
  \bibfield  {author} {\bibinfo {author} {\bibfnamefont {W.~D.}\ \bibnamefont
  {Goldberger}}\ and\ \bibinfo {author} {\bibfnamefont {M.~B.}\ \bibnamefont
  {Wise}},\ }\href {\doibase 10.1016/S0370-2693(00)00099-X} {\bibfield
  {journal} {\bibinfo  {journal} {Phys. Lett. B}\ }\textbf {\bibinfo {volume}
  {475}},\ \bibinfo {pages} {275} (\bibinfo {year} {2000})},\ \Eprint
  {http://arxiv.org/abs/hep-ph/9911457} {arXiv:hep-ph/9911457} \BibitemShut
  {NoStop}%
\bibitem [{\citenamefont {Csaki}\ \emph {et~al.}(2001)\citenamefont {Csaki},
  \citenamefont {Graesser},\ and\ \citenamefont {Kribs}}]{Csaki:2000zn}%
  \BibitemOpen
  \bibfield  {author} {\bibinfo {author} {\bibfnamefont {C.}~\bibnamefont
  {Csaki}}, \bibinfo {author} {\bibfnamefont {M.~L.}\ \bibnamefont {Graesser}},
  \ and\ \bibinfo {author} {\bibfnamefont {G.~D.}\ \bibnamefont {Kribs}},\
  }\href {\doibase 10.1103/PhysRevD.63.065002} {\bibfield  {journal} {\bibinfo
  {journal} {Phys. Rev. D}\ }\textbf {\bibinfo {volume} {63}},\ \bibinfo
  {pages} {065002} (\bibinfo {year} {2001})},\ \Eprint
  {http://arxiv.org/abs/hep-th/0008151} {arXiv:hep-th/0008151} \BibitemShut
  {NoStop}%
\bibitem [{\citenamefont {Csaki}\ \emph {et~al.}(2007)\citenamefont {Csaki},
  \citenamefont {Hubisz},\ and\ \citenamefont {Lee}}]{Csaki:2007ns}%
  \BibitemOpen
  \bibfield  {author} {\bibinfo {author} {\bibfnamefont {C.}~\bibnamefont
  {Csaki}}, \bibinfo {author} {\bibfnamefont {J.}~\bibnamefont {Hubisz}}, \
  and\ \bibinfo {author} {\bibfnamefont {S.~J.}\ \bibnamefont {Lee}},\ }\href
  {\doibase 10.1103/PhysRevD.76.125015} {\bibfield  {journal} {\bibinfo
  {journal} {Phys. Rev. D}\ }\textbf {\bibinfo {volume} {76}},\ \bibinfo
  {pages} {125015} (\bibinfo {year} {2007})},\ \Eprint
  {http://arxiv.org/abs/0705.3844} {arXiv:0705.3844 [hep-ph]} \BibitemShut
  {NoStop}%
\bibitem [{\citenamefont {Rattazzi}\ and\ \citenamefont
  {Zaffaroni}(2001)}]{Rattazzi:2000hs}%
  \BibitemOpen
  \bibfield  {author} {\bibinfo {author} {\bibfnamefont {R.}~\bibnamefont
  {Rattazzi}}\ and\ \bibinfo {author} {\bibfnamefont {A.}~\bibnamefont
  {Zaffaroni}},\ }\href {\doibase 10.1088/1126-6708/2001/04/021} {\bibfield
  {journal} {\bibinfo  {journal} {JHEP}\ }\textbf {\bibinfo {volume} {04}},\
  \bibinfo {pages} {021} (\bibinfo {year} {2001})},\ \Eprint
  {http://arxiv.org/abs/hep-th/0012248} {arXiv:hep-th/0012248} \BibitemShut
  {NoStop}%
\bibitem [{\citenamefont {Arkani-Hamed}\ \emph {et~al.}(2001)\citenamefont
  {Arkani-Hamed}, \citenamefont {Porrati},\ and\ \citenamefont
  {Randall}}]{Arkani-Hamed:2000ijo}%
  \BibitemOpen
  \bibfield  {author} {\bibinfo {author} {\bibfnamefont {N.}~\bibnamefont
  {Arkani-Hamed}}, \bibinfo {author} {\bibfnamefont {M.}~\bibnamefont
  {Porrati}}, \ and\ \bibinfo {author} {\bibfnamefont {L.}~\bibnamefont
  {Randall}},\ }\href {\doibase 10.1088/1126-6708/2001/08/017} {\bibfield
  {journal} {\bibinfo  {journal} {JHEP}\ }\textbf {\bibinfo {volume} {08}},\
  \bibinfo {pages} {017} (\bibinfo {year} {2001})},\ \Eprint
  {http://arxiv.org/abs/hep-th/0012148} {arXiv:hep-th/0012148} \BibitemShut
  {NoStop}%
\bibitem [{\citenamefont {Bellazzini}\ \emph {et~al.}(2013)\citenamefont
  {Bellazzini}, \citenamefont {Csaki}, \citenamefont {Hubisz}, \citenamefont
  {Serra},\ and\ \citenamefont {Terning}}]{Bellazzini:2012vz}%
  \BibitemOpen
  \bibfield  {author} {\bibinfo {author} {\bibfnamefont {B.}~\bibnamefont
  {Bellazzini}}, \bibinfo {author} {\bibfnamefont {C.}~\bibnamefont {Csaki}},
  \bibinfo {author} {\bibfnamefont {J.}~\bibnamefont {Hubisz}}, \bibinfo
  {author} {\bibfnamefont {J.}~\bibnamefont {Serra}}, \ and\ \bibinfo {author}
  {\bibfnamefont {J.}~\bibnamefont {Terning}},\ }\href {\doibase
  10.1140/epjc/s10052-013-2333-x} {\bibfield  {journal} {\bibinfo  {journal}
  {Eur. Phys. J. C}\ }\textbf {\bibinfo {volume} {73}},\ \bibinfo {pages}
  {2333} (\bibinfo {year} {2013})},\ \Eprint {http://arxiv.org/abs/1209.3299}
  {arXiv:1209.3299 [hep-ph]} \BibitemShut {NoStop}%
\bibitem [{\citenamefont {Chacko}\ and\ \citenamefont
  {Mishra}(2013)}]{Chacko:2012sy}%
  \BibitemOpen
  \bibfield  {author} {\bibinfo {author} {\bibfnamefont {Z.}~\bibnamefont
  {Chacko}}\ and\ \bibinfo {author} {\bibfnamefont {R.~K.}\ \bibnamefont
  {Mishra}},\ }\href {\doibase 10.1103/PhysRevD.87.115006} {\bibfield
  {journal} {\bibinfo  {journal} {Phys. Rev. D}\ }\textbf {\bibinfo {volume}
  {87}},\ \bibinfo {pages} {115006} (\bibinfo {year} {2013})},\ \Eprint
  {http://arxiv.org/abs/1209.3022} {arXiv:1209.3022 [hep-ph]} \BibitemShut
  {NoStop}%
\bibitem [{\citenamefont {Fubini}(1976)}]{Fubini:1976jm}%
  \BibitemOpen
  \bibfield  {author} {\bibinfo {author} {\bibfnamefont {S.}~\bibnamefont
  {Fubini}},\ }\href {\doibase 10.1007/BF02785664} {\bibfield  {journal}
  {\bibinfo  {journal} {Nuovo Cim. A}\ }\textbf {\bibinfo {volume} {34}},\
  \bibinfo {pages} {521} (\bibinfo {year} {1976})}\BibitemShut {NoStop}%
\bibitem [{\citenamefont {Contino}\ \emph {et~al.}(2010)\citenamefont
  {Contino}, \citenamefont {Pomarol},\ and\ \citenamefont
  {Rattazzi}}]{CPRtalk1}%
  \BibitemOpen
  \bibfield  {author} {\bibinfo {author} {\bibfnamefont {R.}~\bibnamefont
  {Contino}}, \bibinfo {author} {\bibfnamefont {A.}~\bibnamefont {Pomarol}}, \
  and\ \bibinfo {author} {\bibfnamefont {R.}~\bibnamefont {Rattazzi}},\
  }\href@noop {} {}\bibinfo {howpublished} {talk by {R.~Rattazzi} at {Planck}
  2010, {CERN}
  \href{https://indico.cern.ch/event/75810/contributions/1250635/attachments/1050757/1498158/Rattazzi.pdf}{[slides]}}
  (\bibinfo {year} {2010})\BibitemShut {NoStop}%
\bibitem [{CPR(2010)}]{CPRtalk2}%
  \BibitemOpen
  \href@noop {} {}\bibinfo {howpublished} {talk by {A.~Pomarol} at 2010
  {Madrid} {Christmas} {Workshop}
  \href{https://www.ift.uam-csic.es/www2/workshops/Xmas10/doc/pomarol.pdf}{[slides]}}
  (\bibinfo {year} {2010})\BibitemShut {NoStop}%
\bibitem [{\citenamefont {Bellazzini}\ \emph
  {et~al.}(2014{\natexlab{b}})\citenamefont {Bellazzini}, \citenamefont
  {Csaki}, \citenamefont {Hubisz}, \citenamefont {Serra},\ and\ \citenamefont
  {Terning}}]{Bellazzini:2013fga}%
  \BibitemOpen
  \bibfield  {author} {\bibinfo {author} {\bibfnamefont {B.}~\bibnamefont
  {Bellazzini}}, \bibinfo {author} {\bibfnamefont {C.}~\bibnamefont {Csaki}},
  \bibinfo {author} {\bibfnamefont {J.}~\bibnamefont {Hubisz}}, \bibinfo
  {author} {\bibfnamefont {J.}~\bibnamefont {Serra}}, \ and\ \bibinfo {author}
  {\bibfnamefont {J.}~\bibnamefont {Terning}},\ }\href {\doibase
  10.1140/epjc/s10052-014-2790-x} {\bibfield  {journal} {\bibinfo  {journal}
  {Eur. Phys. J. C}\ }\textbf {\bibinfo {volume} {74}},\ \bibinfo {pages}
  {2790} (\bibinfo {year} {2014}{\natexlab{b}})},\ \Eprint
  {http://arxiv.org/abs/1305.3919} {arXiv:1305.3919 [hep-th]} \BibitemShut
  {NoStop}%
\bibitem [{\citenamefont {Komargodski}\ and\ \citenamefont
  {Schwimmer}(2011)}]{Komargodski:2011vj}%
  \BibitemOpen
  \bibfield  {author} {\bibinfo {author} {\bibfnamefont {Z.}~\bibnamefont
  {Komargodski}}\ and\ \bibinfo {author} {\bibfnamefont {A.}~\bibnamefont
  {Schwimmer}},\ }\href {\doibase 10.1007/JHEP12(2011)099} {\bibfield
  {journal} {\bibinfo  {journal} {JHEP}\ }\textbf {\bibinfo {volume} {12}},\
  \bibinfo {pages} {099} (\bibinfo {year} {2011})},\ \Eprint
  {http://arxiv.org/abs/1107.3987} {arXiv:1107.3987 [hep-th]} \BibitemShut
  {NoStop}%
\bibitem [{\citenamefont {Henningson}\ and\ \citenamefont
  {Skenderis}(1998)}]{Henningson:1998gx}%
  \BibitemOpen
  \bibfield  {author} {\bibinfo {author} {\bibfnamefont {M.}~\bibnamefont
  {Henningson}}\ and\ \bibinfo {author} {\bibfnamefont {K.}~\bibnamefont
  {Skenderis}},\ }\href {\doibase 10.1088/1126-6708/1998/07/023} {\bibfield
  {journal} {\bibinfo  {journal} {JHEP}\ }\textbf {\bibinfo {volume} {07}},\
  \bibinfo {pages} {023} (\bibinfo {year} {1998})},\ \Eprint
  {http://arxiv.org/abs/hep-th/9806087} {arXiv:hep-th/9806087} \BibitemShut
  {NoStop}%
\bibitem [{\citenamefont {Nojiri}\ and\ \citenamefont
  {Odintsov}(2000)}]{Nojiri:1999mh}%
  \BibitemOpen
  \bibfield  {author} {\bibinfo {author} {\bibfnamefont {S.}~\bibnamefont
  {Nojiri}}\ and\ \bibinfo {author} {\bibfnamefont {S.~D.}\ \bibnamefont
  {Odintsov}},\ }\href {\doibase 10.1142/S0217751X00000197} {\bibfield
  {journal} {\bibinfo  {journal} {Int. J. Mod. Phys. A}\ }\textbf {\bibinfo
  {volume} {15}},\ \bibinfo {pages} {413} (\bibinfo {year} {2000})},\ \Eprint
  {http://arxiv.org/abs/hep-th/9903033} {arXiv:hep-th/9903033} \BibitemShut
  {NoStop}%
\bibitem [{\citenamefont {Freedman}\ \emph {et~al.}(1999)\citenamefont
  {Freedman}, \citenamefont {Gubser}, \citenamefont {Pilch},\ and\
  \citenamefont {Warner}}]{Freedman:1999gp}%
  \BibitemOpen
  \bibfield  {author} {\bibinfo {author} {\bibfnamefont {D.~Z.}\ \bibnamefont
  {Freedman}}, \bibinfo {author} {\bibfnamefont {S.~S.}\ \bibnamefont
  {Gubser}}, \bibinfo {author} {\bibfnamefont {K.}~\bibnamefont {Pilch}}, \
  and\ \bibinfo {author} {\bibfnamefont {N.~P.}\ \bibnamefont {Warner}},\
  }\href {\doibase 10.4310/ATMP.1999.v3.n2.a7} {\bibfield  {journal} {\bibinfo
  {journal} {Adv. Theor. Math. Phys.}\ }\textbf {\bibinfo {volume} {3}},\
  \bibinfo {pages} {363} (\bibinfo {year} {1999})},\ \Eprint
  {http://arxiv.org/abs/hep-th/9904017} {arXiv:hep-th/9904017} \BibitemShut
  {NoStop}%
\bibitem [{\citenamefont {Girardello}\ \emph {et~al.}(2000)\citenamefont
  {Girardello}, \citenamefont {Petrini}, \citenamefont {Porrati},\ and\
  \citenamefont {Zaffaroni}}]{Girardello:1999bd}%
  \BibitemOpen
  \bibfield  {author} {\bibinfo {author} {\bibfnamefont {L.}~\bibnamefont
  {Girardello}}, \bibinfo {author} {\bibfnamefont {M.}~\bibnamefont {Petrini}},
  \bibinfo {author} {\bibfnamefont {M.}~\bibnamefont {Porrati}}, \ and\
  \bibinfo {author} {\bibfnamefont {A.}~\bibnamefont {Zaffaroni}},\ }\href
  {\doibase 10.1016/S0550-3213(99)00764-6} {\bibfield  {journal} {\bibinfo
  {journal} {Nucl. Phys. B}\ }\textbf {\bibinfo {volume} {569}},\ \bibinfo
  {pages} {451} (\bibinfo {year} {2000})},\ \Eprint
  {http://arxiv.org/abs/hep-th/9909047} {arXiv:hep-th/9909047} \BibitemShut
  {NoStop}%
\bibitem [{\citenamefont {Imbimbo}\ \emph {et~al.}(2000)\citenamefont
  {Imbimbo}, \citenamefont {Schwimmer}, \citenamefont {Theisen},\ and\
  \citenamefont {Yankielowicz}}]{Imbimbo:1999bj}%
  \BibitemOpen
  \bibfield  {author} {\bibinfo {author} {\bibfnamefont {C.}~\bibnamefont
  {Imbimbo}}, \bibinfo {author} {\bibfnamefont {A.}~\bibnamefont {Schwimmer}},
  \bibinfo {author} {\bibfnamefont {S.}~\bibnamefont {Theisen}}, \ and\
  \bibinfo {author} {\bibfnamefont {S.}~\bibnamefont {Yankielowicz}},\ }\href
  {\doibase 10.1088/0264-9381/17/5/322} {\bibfield  {journal} {\bibinfo
  {journal} {Class. Quant. Grav.}\ }\textbf {\bibinfo {volume} {17}},\ \bibinfo
  {pages} {1129} (\bibinfo {year} {2000})},\ \Eprint
  {http://arxiv.org/abs/hep-th/9910267} {arXiv:hep-th/9910267} \BibitemShut
  {NoStop}%
\bibitem [{\citenamefont {Myers}\ and\ \citenamefont
  {Sinha}(2011)}]{Myers:2010tj}%
  \BibitemOpen
  \bibfield  {author} {\bibinfo {author} {\bibfnamefont {R.~C.}\ \bibnamefont
  {Myers}}\ and\ \bibinfo {author} {\bibfnamefont {A.}~\bibnamefont {Sinha}},\
  }\href {\doibase 10.1007/JHEP01(2011)125} {\bibfield  {journal} {\bibinfo
  {journal} {JHEP}\ }\textbf {\bibinfo {volume} {01}},\ \bibinfo {pages} {125}
  (\bibinfo {year} {2011})},\ \Eprint {http://arxiv.org/abs/1011.5819}
  {arXiv:1011.5819 [hep-th]} \BibitemShut {NoStop}%
\bibitem [{\citenamefont {Low}\ and\ \citenamefont
  {Manohar}(2002)}]{Low:2001bw}%
  \BibitemOpen
  \bibfield  {author} {\bibinfo {author} {\bibfnamefont {I.}~\bibnamefont
  {Low}}\ and\ \bibinfo {author} {\bibfnamefont {A.~V.}\ \bibnamefont
  {Manohar}},\ }\href {\doibase 10.1103/PhysRevLett.88.101602} {\bibfield
  {journal} {\bibinfo  {journal} {Phys. Rev. Lett.}\ }\textbf {\bibinfo
  {volume} {88}},\ \bibinfo {pages} {101602} (\bibinfo {year} {2002})},\
  \Eprint {http://arxiv.org/abs/hep-th/0110285} {arXiv:hep-th/0110285}
  \BibitemShut {NoStop}%
\bibitem [{\citenamefont {Volkov}(1973)}]{Volkov:1973vd}%
  \BibitemOpen
  \bibfield  {author} {\bibinfo {author} {\bibfnamefont {D.~V.}\ \bibnamefont
  {Volkov}},\ }\href@noop {} {\bibfield  {journal} {\bibinfo  {journal} {Fiz.
  Elem. Chast. Atom. Yadra}\ }\textbf {\bibinfo {volume} {4}},\ \bibinfo
  {pages} {3} (\bibinfo {year} {1973})}\BibitemShut {NoStop}%
\bibitem [{\citenamefont {Duff}(1994)}]{Duff:1993wm}%
  \BibitemOpen
  \bibfield  {author} {\bibinfo {author} {\bibfnamefont {M.~J.}\ \bibnamefont
  {Duff}},\ }\href {\doibase 10.1088/0264-9381/11/6/004} {\bibfield  {journal}
  {\bibinfo  {journal} {Class. Quant. Grav.}\ }\textbf {\bibinfo {volume}
  {11}},\ \bibinfo {pages} {1387} (\bibinfo {year} {1994})},\ \Eprint
  {http://arxiv.org/abs/hep-th/9308075} {arXiv:hep-th/9308075} \BibitemShut
  {NoStop}%
\bibitem [{\citenamefont {Schwimmer}\ and\ \citenamefont
  {Theisen}(2011)}]{Schwimmer:2010za}%
  \BibitemOpen
  \bibfield  {author} {\bibinfo {author} {\bibfnamefont {A.}~\bibnamefont
  {Schwimmer}}\ and\ \bibinfo {author} {\bibfnamefont {S.}~\bibnamefont
  {Theisen}},\ }\href {\doibase 10.1016/j.nuclphysb.2011.02.003} {\bibfield
  {journal} {\bibinfo  {journal} {Nucl. Phys. B}\ }\textbf {\bibinfo {volume}
  {847}},\ \bibinfo {pages} {590} (\bibinfo {year} {2011})},\ \Eprint
  {http://arxiv.org/abs/1011.0696} {arXiv:1011.0696 [hep-th]} \BibitemShut
  {NoStop}%
\bibitem [{\citenamefont {Fradkin}\ and\ \citenamefont
  {Tseytlin}(1984)}]{Fradkin:1983tg}%
  \BibitemOpen
  \bibfield  {author} {\bibinfo {author} {\bibfnamefont {E.~S.}\ \bibnamefont
  {Fradkin}}\ and\ \bibinfo {author} {\bibfnamefont {A.~A.}\ \bibnamefont
  {Tseytlin}},\ }\href {\doibase 10.1016/0370-2693(84)90668-3} {\bibfield
  {journal} {\bibinfo  {journal} {Phys. Lett. B}\ }\textbf {\bibinfo {volume}
  {134}},\ \bibinfo {pages} {187} (\bibinfo {year} {1984})}\BibitemShut
  {NoStop}%
\bibitem [{\citenamefont {Riegert}(1984)}]{Riegert:1984kt}%
  \BibitemOpen
  \bibfield  {author} {\bibinfo {author} {\bibfnamefont {R.~J.}\ \bibnamefont
  {Riegert}},\ }\href {\doibase 10.1016/0370-2693(84)90983-3} {\bibfield
  {journal} {\bibinfo  {journal} {Phys. Lett. B}\ }\textbf {\bibinfo {volume}
  {134}},\ \bibinfo {pages} {56} (\bibinfo {year} {1984})}\BibitemShut
  {NoStop}%
\bibitem [{\citenamefont {von Harling}\ and\ \citenamefont
  {Servant}(2018)}]{vonHarling:2017yew}%
  \BibitemOpen
  \bibfield  {author} {\bibinfo {author} {\bibfnamefont {B.}~\bibnamefont {von
  Harling}}\ and\ \bibinfo {author} {\bibfnamefont {G.}~\bibnamefont
  {Servant}},\ }\href {\doibase 10.1007/JHEP01(2018)159} {\bibfield  {journal}
  {\bibinfo  {journal} {JHEP}\ }\textbf {\bibinfo {volume} {01}},\ \bibinfo
  {pages} {159} (\bibinfo {year} {2018})},\ \Eprint
  {http://arxiv.org/abs/1711.11554} {arXiv:1711.11554 [hep-ph]} \BibitemShut
  {NoStop}%
\bibitem [{\citenamefont {Myers}\ \emph {et~al.}(2010)\citenamefont {Myers},
  \citenamefont {Paulos},\ and\ \citenamefont {Sinha}}]{Myers:2010jv}%
  \BibitemOpen
  \bibfield  {author} {\bibinfo {author} {\bibfnamefont {R.~C.}\ \bibnamefont
  {Myers}}, \bibinfo {author} {\bibfnamefont {M.~F.}\ \bibnamefont {Paulos}}, \
  and\ \bibinfo {author} {\bibfnamefont {A.}~\bibnamefont {Sinha}},\ }\href
  {\doibase 10.1007/JHEP08(2010)035} {\bibfield  {journal} {\bibinfo  {journal}
  {JHEP}\ }\textbf {\bibinfo {volume} {08}},\ \bibinfo {pages} {035} (\bibinfo
  {year} {2010})},\ \Eprint {http://arxiv.org/abs/1004.2055} {arXiv:1004.2055
  [hep-th]} \BibitemShut {NoStop}%
\bibitem [{\citenamefont {Sundrum}(2003)}]{Sundrum:2003yt}%
  \BibitemOpen
  \bibfield  {author} {\bibinfo {author} {\bibfnamefont {R.}~\bibnamefont
  {Sundrum}},\ }\href@noop {} {\  (\bibinfo {year} {2003})},\ \Eprint
  {http://arxiv.org/abs/hep-th/0312212} {arXiv:hep-th/0312212} \BibitemShut
  {NoStop}%
\bibitem [{\citenamefont {Chamblin}\ \emph {et~al.}(2000)\citenamefont
  {Chamblin}, \citenamefont {Hawking},\ and\ \citenamefont
  {Reall}}]{Chamblin:1999by}%
  \BibitemOpen
  \bibfield  {author} {\bibinfo {author} {\bibfnamefont {A.}~\bibnamefont
  {Chamblin}}, \bibinfo {author} {\bibfnamefont {S.~W.}\ \bibnamefont
  {Hawking}}, \ and\ \bibinfo {author} {\bibfnamefont {H.~S.}\ \bibnamefont
  {Reall}},\ }\href {\doibase 10.1103/PhysRevD.61.065007} {\bibfield  {journal}
  {\bibinfo  {journal} {Phys. Rev. D}\ }\textbf {\bibinfo {volume} {61}},\
  \bibinfo {pages} {065007} (\bibinfo {year} {2000})},\ \Eprint
  {http://arxiv.org/abs/hep-th/9909205} {arXiv:hep-th/9909205} \BibitemShut
  {NoStop}%
\bibitem [{\citenamefont {Emparan}\ \emph
  {et~al.}(2000{\natexlab{a}})\citenamefont {Emparan}, \citenamefont
  {Horowitz},\ and\ \citenamefont {Myers}}]{Emparan:1999wa}%
  \BibitemOpen
  \bibfield  {author} {\bibinfo {author} {\bibfnamefont {R.}~\bibnamefont
  {Emparan}}, \bibinfo {author} {\bibfnamefont {G.~T.}\ \bibnamefont
  {Horowitz}}, \ and\ \bibinfo {author} {\bibfnamefont {R.~C.}\ \bibnamefont
  {Myers}},\ }\href {\doibase 10.1088/1126-6708/2000/01/007} {\bibfield
  {journal} {\bibinfo  {journal} {JHEP}\ }\textbf {\bibinfo {volume} {01}},\
  \bibinfo {pages} {007} (\bibinfo {year} {2000}{\natexlab{a}})},\ \Eprint
  {http://arxiv.org/abs/hep-th/9911043} {arXiv:hep-th/9911043} \BibitemShut
  {NoStop}%
\bibitem [{\citenamefont {Emparan}\ \emph
  {et~al.}(2000{\natexlab{b}})\citenamefont {Emparan}, \citenamefont
  {Horowitz},\ and\ \citenamefont {Myers}}]{Emparan:1999fd}%
  \BibitemOpen
  \bibfield  {author} {\bibinfo {author} {\bibfnamefont {R.}~\bibnamefont
  {Emparan}}, \bibinfo {author} {\bibfnamefont {G.~T.}\ \bibnamefont
  {Horowitz}}, \ and\ \bibinfo {author} {\bibfnamefont {R.~C.}\ \bibnamefont
  {Myers}},\ }\href {\doibase 10.1088/1126-6708/2000/01/021} {\bibfield
  {journal} {\bibinfo  {journal} {JHEP}\ }\textbf {\bibinfo {volume} {01}},\
  \bibinfo {pages} {021} (\bibinfo {year} {2000}{\natexlab{b}})},\ \Eprint
  {http://arxiv.org/abs/hep-th/9912135} {arXiv:hep-th/9912135} \BibitemShut
  {NoStop}%
\bibitem [{\citenamefont {Giddings}\ and\ \citenamefont
  {Katz}(2001)}]{Giddings:2000ay}%
  \BibitemOpen
  \bibfield  {author} {\bibinfo {author} {\bibfnamefont {S.~B.}\ \bibnamefont
  {Giddings}}\ and\ \bibinfo {author} {\bibfnamefont {E.}~\bibnamefont
  {Katz}},\ }\href {\doibase 10.1063/1.1377036} {\bibfield  {journal} {\bibinfo
   {journal} {J. Math. Phys.}\ }\textbf {\bibinfo {volume} {42}},\ \bibinfo
  {pages} {3082} (\bibinfo {year} {2001})},\ \Eprint
  {http://arxiv.org/abs/hep-th/0009176} {arXiv:hep-th/0009176} \BibitemShut
  {NoStop}%
\bibitem [{\citenamefont {Emparan}\ \emph {et~al.}(2002)\citenamefont
  {Emparan}, \citenamefont {Fabbri},\ and\ \citenamefont
  {Kaloper}}]{Emparan:2002px}%
  \BibitemOpen
  \bibfield  {author} {\bibinfo {author} {\bibfnamefont {R.}~\bibnamefont
  {Emparan}}, \bibinfo {author} {\bibfnamefont {A.}~\bibnamefont {Fabbri}}, \
  and\ \bibinfo {author} {\bibfnamefont {N.}~\bibnamefont {Kaloper}},\ }\href
  {\doibase 10.1088/1126-6708/2002/08/043} {\bibfield  {journal} {\bibinfo
  {journal} {JHEP}\ }\textbf {\bibinfo {volume} {08}},\ \bibinfo {pages} {043}
  (\bibinfo {year} {2002})},\ \Eprint {http://arxiv.org/abs/hep-th/0206155}
  {arXiv:hep-th/0206155} \BibitemShut {NoStop}%
\bibitem [{\citenamefont {Meade}\ and\ \citenamefont
  {Randall}(2008)}]{Meade:2007sz}%
  \BibitemOpen
  \bibfield  {author} {\bibinfo {author} {\bibfnamefont {P.}~\bibnamefont
  {Meade}}\ and\ \bibinfo {author} {\bibfnamefont {L.}~\bibnamefont
  {Randall}},\ }\href {\doibase 10.1088/1126-6708/2008/05/003} {\bibfield
  {journal} {\bibinfo  {journal} {JHEP}\ }\textbf {\bibinfo {volume} {05}},\
  \bibinfo {pages} {003} (\bibinfo {year} {2008})},\ \Eprint
  {http://arxiv.org/abs/0708.3017} {arXiv:0708.3017 [hep-ph]} \BibitemShut
  {NoStop}%
\bibitem [{\citenamefont {Csaki}\ \emph
  {et~al.}(2000{\natexlab{b}})\citenamefont {Csaki}, \citenamefont {Erlich},
  \citenamefont {Grojean},\ and\ \citenamefont {Hollowood}}]{Csaki:2000wz}%
  \BibitemOpen
  \bibfield  {author} {\bibinfo {author} {\bibfnamefont {C.}~\bibnamefont
  {Csaki}}, \bibinfo {author} {\bibfnamefont {J.}~\bibnamefont {Erlich}},
  \bibinfo {author} {\bibfnamefont {C.}~\bibnamefont {Grojean}}, \ and\
  \bibinfo {author} {\bibfnamefont {T.~J.}\ \bibnamefont {Hollowood}},\ }\href
  {\doibase 10.1016/S0550-3213(00)00390-4} {\bibfield  {journal} {\bibinfo
  {journal} {Nucl. Phys. B}\ }\textbf {\bibinfo {volume} {584}},\ \bibinfo
  {pages} {359} (\bibinfo {year} {2000}{\natexlab{b}})},\ \Eprint
  {http://arxiv.org/abs/hep-th/0004133} {arXiv:hep-th/0004133} \BibitemShut
  {NoStop}%
\bibitem [{\citenamefont {Cheung}\ \emph {et~al.}(2018)\citenamefont {Cheung},
  \citenamefont {Liu},\ and\ \citenamefont {Remmen}}]{Cheung:2018cwt}%
  \BibitemOpen
  \bibfield  {author} {\bibinfo {author} {\bibfnamefont {C.}~\bibnamefont
  {Cheung}}, \bibinfo {author} {\bibfnamefont {J.}~\bibnamefont {Liu}}, \ and\
  \bibinfo {author} {\bibfnamefont {G.~N.}\ \bibnamefont {Remmen}},\ }\href
  {\doibase 10.1007/JHEP10(2018)004} {\bibfield  {journal} {\bibinfo  {journal}
  {JHEP}\ }\textbf {\bibinfo {volume} {10}},\ \bibinfo {pages} {004} (\bibinfo
  {year} {2018})},\ \Eprint {http://arxiv.org/abs/1801.08546} {arXiv:1801.08546
  [hep-th]} \BibitemShut {NoStop}%
\bibitem [{\citenamefont {Liu}\ and\ \citenamefont {Sabra}(2010)}]{Liu:2008zf}%
  \BibitemOpen
  \bibfield  {author} {\bibinfo {author} {\bibfnamefont {J.~T.}\ \bibnamefont
  {Liu}}\ and\ \bibinfo {author} {\bibfnamefont {W.~A.}\ \bibnamefont
  {Sabra}},\ }\href {\doibase 10.1088/0264-9381/27/17/175014} {\bibfield
  {journal} {\bibinfo  {journal} {Class. Quant. Grav.}\ }\textbf {\bibinfo
  {volume} {27}},\ \bibinfo {pages} {175014} (\bibinfo {year} {2010})},\
  \Eprint {http://arxiv.org/abs/0807.1256} {arXiv:0807.1256 [hep-th]}
  \BibitemShut {NoStop}%
\end{thebibliography}%
\end{document}